\documentclass[aps,prd,superscriptaddress,showpacs,showkeys,nofootinbib,nobibnotes,twocolumn,preprintnumbers,floatfix]{revtex4}

\usepackage{amsmath}
\usepackage{amssymb}
\usepackage{graphicx}
\usepackage{color}[]
\usepackage{bm}  
 
\newcommand{\del}{\partial}

\begin{document}

\preprint{ADP-12-09/T776}

\title{$\Delta I=1/2$ rule for kaon decays derived from QCD infrared fixed point}
\author{R.~J.~Crewther}
\email{rcrewthe@physics.adelaide.edu.au}
\affiliation{CSSM and ARC Centre of Excellence for Particle Physics at 
the Tera-scale, Department of Physics, University of Adelaide, 
Adelaide, South Australia 5005, Australia} 
\author{Lewis~C.~Tunstall}
\email{tunstall@itp.unibe.ch}
\affiliation{CSSM and ARC Centre of Excellence for Particle Physics at 
the Tera-scale, Department of Physics, University of Adelaide, 
Adelaide, South Australia 5005, Australia} 
\affiliation{Berkeley Center for Theoretical Physics, Department of Physics, 
University of California, Berkeley, California 94720, USA} 
\affiliation{Albert Einstein Centre for Fundamental Physics, 
Institute for Theoretical Physics, University of Bern, Sidlerstrasse 5, 
3012 Bern, Switzerland}

\begin{abstract}
This article gives details of our proposal to replace ordinary chiral 
$SU(3)_L\times SU(3)_R$ perturbation theory $\chi$PT$_3$ by three-flavor 
chiral-scale perturbation theory $\chi$PT$_\sigma$. In $\chi$PT$_\sigma$, 
amplitudes are expanded at low energies and small $u,d,s$ quark masses
about an infrared fixed point $\alpha^{}_\mathrm{IR}$ of three-flavor QCD.  
At $\alpha^{}_\mathrm{IR}$, the quark condensate 
$\langle \bar{q}q\rangle_{\mathrm{vac}} \not= 0$ induces \emph{nine} 
Nambu-Goldstone bosons: $\pi, K, \eta$ and a $0^{++}$ QCD dilaton 
$\sigma$. Physically, $\sigma$ appears as the $f_{0}(500)$ resonance,
a pole at a complex mass with real part $\lesssim m_K$. The 
$\Delta I=1/2$ rule for nonleptonic $K$-decays is then a 
\emph{consequence} of $\chi$PT$_\sigma$, with a $K_S\sigma$ coupling 
fixed by data for $\gamma\gamma\rightarrow\pi\pi$ and 
$K_{S} \to \gamma\gamma$. We estimate $R_\mathrm{IR} \approx 5$ 
for the nonperturbative Drell-Yan ratio 
$R = \sigma(e^{+}e^{-}\rightarrow\mathrm{hadrons})/ 
   \sigma(e^{+}e^{-}\rightarrow\mu^{+}\mu^{-})$ at $\alpha^{}_\mathrm{IR}$  
and show that, in the many-color limit, $\sigma/f_0$ becomes a narrow 
$q\bar{q}$ state with planar-gluon corrections. Rules for the order of 
terms in $\chi$PT$_\sigma$ loop expansions are derived in Appendix \ref{AppA}  
and extended in Appendix \ref{AppB} to include inverse-power Li-Pagels 
singularities due to external operators. This relates to an observation 
that, for $\gamma\gamma$ channels, partial conservation of the dilatation 
current is \emph{not} equivalent to $\sigma$-pole dominance.
\end{abstract}



\maketitle

\section{Summary}
\label{Intro}
\begin{figure}[t]
\center\includegraphics[scale=0.75]{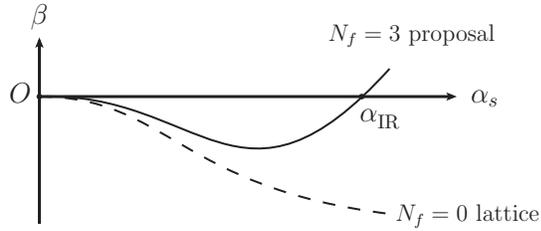}
\caption{The solid line shows a three-flavor $\beta$ function (or
better, a QCD version \cite{Grun82} of the Gell-Mann--Low $\Psi$ 
function) with an infrared fixed point $\alpha^{}_{\mathrm{IR}}$
at which $\alpha_s$ freezes \cite{pinch,Steve,freeze,Brod,lattice,DSE} 
but the manifest scale invariance of \cite{CBZ,Appel,Del10} is \emph{avoided}.
The existence of $\alpha^{}_{\mathrm{IR}}$ for small $N_f$ values is not 
entirely settled. The dashed line shows the original lattice result 
\cite{Lusc94} for $N_f = 0$ (no quarks) where $\beta$ remains negative and 
becomes linear at large $\alpha_s$.}  
\label{fig:beta}
\end{figure}

The precise determination of the mass and width of the $f_0(500)$ resonance 
\cite{Cap06,Kam11, PDG} prompted us \cite{us} to revisit an old idea 
\cite{RJC70,Ell70} that the chiral condensate 
$\langle \bar{q}q \rangle_{\mathrm{vac}} \not= 0$ may also be a condensate 
for scale transformations in the chiral $SU(3)_L\times SU(3)_R$ limit. This 
may occur in QCD if the heavy quarks $t,b,c$ are first decoupled and then 
the strong coupling%
\footnotemark[1]\addtocounter{footnote}{1}
\footnotetext[1]{We have $[D_\mu\,,\,D_\nu] 
= ig G^a_{\mu\nu}T^a$ where $D_\mu$ is the covariant derivative, $\{T^a\}$
generate the gauge group, $\alpha_s = g^2/4\pi$ is the strong 
coupling, and $\beta = \mu\del\alpha_{s}/\del\mu$ and 
$\gamma_{m} = \mu\del\ln m_q/\del\mu$ refer to a 
mass-independent renormalization scheme with scale $\mu$.}%
$\alpha_s$ of the resulting three-flavor theory runs nonperturbatively 
to a fixed point $\alpha^{}_{\mathrm{IR}}$ in the infrared limit 
(Fig.\ \ref{fig:beta}). At that point, $\beta(\alpha^{}_{\mathrm{IR}})$ 
vanishes, so the gluonic term in the strong trace anomaly \cite{Mink76}
\begin{equation}\label{eqn:anomaly}
\theta^\mu_\mu
 =\frac{\beta(\alpha_{s})}{4\alpha_{s}} G^a_{\mu\nu}G^{a\mu\nu}
         + \bigl(1 + \gamma_{m}(\alpha_{s})\bigr)\sum_{q=u,d,s} m_{q}\bar{q}q
\end{equation}
is absent, which implies 
\begin{align}  
\left.\theta^\mu_\mu\right|_{\alpha_s = \alpha^{}_{\mathrm{IR}}}
 &= \bigl(1 + \gamma_{m}(\alpha^{}_{\mathrm{IR}})\bigr)
    (m_u\bar{u}u + m_d\bar{d}d + m_s\bar{s}s) \nonumber \\
 &\to 0\ , \ SU(3)_L\times SU(3)_R \mbox{ limit}
\label{scale}\end{align}
and hence a $0^{++}$ QCD dilaton%
\footnotemark[2]\addtocounter{footnote}{1}
\footnotetext[2]{\label{dilaton_def}%
We reserve the term \emph{dilaton} and notation $\sigma$ for a 
NG boson due to scale invariance being preserved by 
the Hamiltonian but broken by the vacuum, in some limit. We are 
\emph{not} talking about the $\sigma$-model, scalar gluonium \cite{glue}, 
or walking gauge theories \cite{Holdom,Aki86,Appel86,Yamawaki}  
where $\beta \approx 0$ near a scale-invariant vacuum \cite{CBZ,Appel,Del10} 
and proposals for ``dilatons'' \cite{Yamawaki,Bando,Appel10} seem unlikely
\cite{HoldomTerning}.}%
$\sigma$ due to quark condensation.%
\footnote{\label{condensate}%
In field and string theory, it is often stated that Green's 
functions are manifestly conformal invariant for $\beta = 0$. This assumes 
that, as in perturbative theories with $\beta = 0$, there are no scale 
condensates. If a scale condensate is present, conformal invariance becomes 
manifest only if \emph{all} four-momenta are spacelike and large.} 
The obvious candidate for this state is the $f_0(500)$, which arises from 
a pole on the second sheet at a complex mass with typical value~\cite{Cap06}
\begin{equation}
m_{f_0} = 441-i\,272 \mbox{ MeV}
\label{f_0}\end{equation} 
and surprisingly small errors \cite{Mink10}.  In all estimates of this type, 
the real part of $m_{f_0}$ is less than $m_K$. 

In Sec.~\ref{Motiv} below, we recall problems with the phenomenology of 
$\chi$PT$_3$ caused by the $f_0$ pole in $0^{++}$ channels, and observe that 
they can be avoided by treating $f_0$ as a Nambu-Goldstone (NG) boson $\sigma$ 
in the limit (\ref{scale}). The result is chiral-scale perturbation theory 
$\chi$PT$_\sigma$, where the NG sector $\{\pi,K,\eta,f_0/\sigma\}$
is clearly separated in scale from other hadrons.

Section~\ref{Lagrangian} introduces the model-independent $\chi$PT$_\sigma$ 
Lagrangian for meson amplitudes expanded in $\alpha_s$ about
$\alpha^{}_{\mathrm{IR}}$ for $m_{u,d,s} \sim 0$. It summarizes soft 
$\pi, K, \eta, \sigma$ meson theorems for three-flavor chiral and scale 
symmetry. For amplitudes where $\sigma$ plays no role, the results agree 
with $\chi$PT$_3$. Results for soft $\sigma$ amplitudes (Sec.~\ref{strong}) 
are similar to those found originally \cite{RJC70,Ell70} but include effects 
due to the gluonic term in (\ref{eqn:anomaly}). In Appendix \ref{AppA}, 
Weinberg's analysis of the $\chi$PT$_2$ loop expansion \cite{Wei79} is 
extended to include $\chi$PT$_\sigma$.

\begin{figure}[t]
\center\includegraphics[scale=.65]{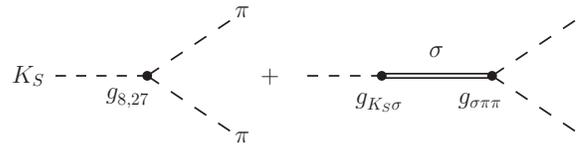}
\caption{Tree diagrams in the effective theory $\chi$PT$_\sigma$ for 
the decay $K_S\to\pi\pi$. The vertex amplitudes due to \textbf{8} and 
\textbf{27} contact couplings $g_8$ and $g_{27}$ are dominated by the 
$\sigma/f_0$ pole amplitude. The magnitude of $g^{}_{K_S\sigma}$ is 
found by applying $\chi$PT$_\sigma$ to $K_S \to \gamma\gamma$ and 
$\gamma\gamma \to \pi\pi$.}
\label{fig:k_pipi}
\end{figure}

Effective electromagnetic and weak operators are then added to simulate 
two-photon processes (Sec.~\ref{Electromag}) and nonleptonic $K$ decays 
(Sec.~\ref{weak_emag}). The main result is a simple explanation of the 
$\Delta I =1/2$ rule for kaon decays:  in the \emph{lowest} order of 
$\chi$PT$_\sigma$, there is a dilaton pole diagram (Fig.\ \ref{fig:k_pipi}) 
which produces most of the $\{\pi\pi\}_{I=0}$ amplitude
\begin{equation}
A_0 = A_{g^{}_{8,27}\,\textrm{vertices}}
+ A_{\sigma\textrm{-pole}} \simeq  A_{\sigma\textrm{-pole}}
\end{equation}
and makes it large relative to the $I=2$ amplitude $A_2$ \cite{PDG}:
\begin{equation}
\bigl|A_0\bigl/A_2\bigr|_\textrm{expt} \simeq 22  \,.
\end{equation} 
We conclude that the ratio of the \textbf{8} and \textbf{27} contact 
couplings $g_8$ and $g_{27}$ is of the order
\begin{equation}
 1 \lesssim \bigl|g_8\bigl/g_{27}\bigr| \lesssim 5
\label{ratio}\end{equation}
indicated by early calculations \cite{Feyn65,Feyn71,Gaill74,Alta74}, and 
\emph{not} the value $22$ found by fitting lowest order $\chi$PT$_3$ to data. 

\begin{figure}[b]
\center\includegraphics[scale=.7]{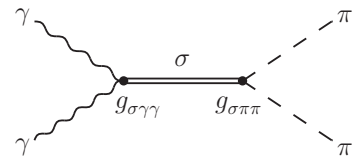}
\caption{Dilaton pole in $\gamma\gamma \to \pi\pi$. In this order 
of $\chi$PT$_\sigma$, diagrams with a $\pi^\pm$ or $K^\pm$ loop 
coupled to both photons must also be included.}
\label{fig:gamgam_pipi}
\end{figure}

In order to obtain a value for the $K_S\sigma$ coupling of 
Fig.~\ref{fig:k_pipi}, we compare the two-photon processes 
$\gamma\gamma \to \pi\pi$ (Fig.~\ref{fig:gamgam_pipi}) and 
$K_S \to \gamma\gamma$. Well-known features of these amplitudes
are the presence of ultraviolet finite $\pi^\pm, K^\pm$ loop 
diagrams coupled to the external photons \cite{DAm86}, and the 
need for a rule \cite{Gasser84,Gasser85,Leut94}
\begin{equation}
A_\mu \sim \del_\mu = O(p)
\label{GLrule}\end{equation} 
specifying the effect of a photon or weak boson field $A_\mu$ 
on the chiral order of terms in loop expansions. These features
are important for our analysis, and in particular, for an
investigation in Sec.~\ref{Electromag} of the relation between
the $\sigma\gamma\gamma$ coupling (Fig.~\ref{fig:gamgam_pipi}) 
and the electromagnetic trace anomaly \cite{RJC72,Ell72} 
\begin{gather}
\widetilde{\theta}^\mu_\mu
= \theta^{\mu}_{\mu} + (R\alpha/6\pi) F_{\mu\nu} F^{\mu\nu} , \notag \\
R=\left.\frac{\sigma(e^{+}e^{-}\rightarrow\mathrm{hadrons})}%
{\sigma(e^{+}e^{-}\rightarrow\mu^{+}\mu^{-})}\right|_{\textrm{energy}\to\infty}
\label{eqn:em_anomaly}
\end{gather}
at the QCD infrared fixed point $\alpha_s = \alpha^{}_\mathrm{IR}$. Here 
$F_{\mu\nu}$ and $\alpha$ are the electromagnetic field strength tensor 
and fine-structure constant, and $\widetilde{\theta}_{\mu\nu}$ is the 
energy-momentum tensor for QCD and QED combined. 

To obtain an approximate result for the decay $\sigma \to \gamma\gamma$, 
the momentum $q$ carried by $\theta^\mu_\mu$ has to be extrapolated from 
$q^2 = 0$ (given exactly by the electromagnetic trace anomaly) to 
$q^2 = m^2_\sigma$. In simple cases, and when photons are absent, this 
amounts to $\sigma$-pole dominance of $\theta^\mu_\mu$, i.e.\  partial 
conservation of the dilatation current (PCDC) \cite{Carr71}, which is the 
direct analogue of partial conservation of the axial-vector current (PCAC) 
for soft-pion amplitudes. However, we find that, unlike PCAC for 
$\pi^0 \to \gamma\gamma$, PCDC for $\sigma \to \gamma\gamma$ is modified 
by meson loop diagrams coupled to photons. In effect, these ultraviolet 
convergent diagrams produce an infrared singularity which is an inverse 
\emph{power} of the light quark mass, arising in the same way as 
conventional Li-Pagels singularities \cite{LiPag,Pagels75}, but sufficiently 
singular to compete with the pole term.

In Appendix \ref{AppB}, we show that, for a fixed number of external operators
coupled purely to the NG sector, these inverse-power singularities do
not upset the convergence of the chiral expansion: relative to the 
corresponding lowest order graph, be it tree or loop, each additional loop  
produces a factor $O(m_q)$ or $O(m_q\ln m_q$). The analysis generalises 
the rule (\ref{GLrule}) for minimal gauge couplings \cite{Gasser84,Gasser85} 
and its extension to axial anomalies \cite{Leut94} to include (a) other 
nonminimal gauge couplings such as the electromagnetic trace anomaly 
(\ref{eqn:em_anomaly}), and (b) external Wilson operators of any kind. 
Appendix \ref{AppC} is a brief note about Eq.~(\ref{eqn:em_anomaly})
for QCD in the physical region $0 < \alpha_s < \alpha^{}_\mathrm{IR}$. 

Unlike other results in this article, our estimate 
\begin{equation}
R^{}_{\mathrm{IR}} \approx 5
\end{equation} 
for the renormalized value of $R$ at the fixed point depends on the 
many-color limit $N_c \to \infty$. This involves the observation 
(Sec.~\ref{Motiv}) that for $N_c$ large, the dilaton 
$\sigma/f_0$ is a  $q\bar{q}$ state, i.e.\ similar to 
$\pi, K, \eta$, but with planar-gluon corrections. Like other 
$q\bar{q}$ resonances, $\sigma/f_0$ has a {\it narrow} 
width in that limit (Sec.~\ref{strong}).

\section{Motivation} 
\label{Motiv}
It may seem odd that new conclusions about QCD can be drawn simply from 
approximate chiral symmetry and $0^{++}$ pole diagrams.  Scalar pole 
dominance for reactions like $K_S \to \pi\pi$ was considered long ago
\cite{Golo80, Volk88, Moro90, Pol02}, it can be easily incorporated 
in a chiral invariant way, and if difficulties with hyperon decays%
\footnote{Accounting for nonleptonic hyperon decays will require either 
$\chi$PT for baryons or the weak sector of the Standard Model to be 
modified.}
are overlooked, theory and experiment for soft $\pi,K,\eta$ amplitudes 
are in excellent agreement, with dispersive corrections included where 
necessary.

The flaw in this picture is contained in another old observation --- 
lowest order $\chi$PT$_3$, if not corrected, typically fails 
for amplitudes which involve both a $0^{++}$ channel and $O(m_K)$ 
extrapolations in momenta: 
\begin{enumerate}
\item Final-state $\pi\pi$ interactions \cite{Tru84} in $K_{\ell 4}$ decays 
      \cite{Tru81} and nonleptonic $K$ \cite{Nev70,Tru88} and $\eta$ 
      \cite{Roi81,Gass85} decays compete with and often dominate 
      purely chiral contributions 
      \cite{Tru84,Tru81,Nev70,Tru88,Roi81,Gass85,Meiss91}.
\item The chiral one-loop prediction for the $K_L \to \pi^0\gamma\gamma$ 
      rate \cite{Eck87} is only 1/3 of the measured value \cite{Kam94}.
\item The lowest order prediction \cite{Bij88,Don88} of a linear 
      rise in the $\gamma\gamma \to \pi^0\pi^0$ cross section disagrees
      \cite{Don93} with the Crystal Ball data \cite{Xal90}.
\end{enumerate}
These facts became evident at a time when it was thought that $0^{++}$ 
resonances below $\approx$ 1 GeV did not exist,%
	\footnote{The $\epsilon(700)$ resonance considered in 
	\cite{RJC70,Ell70,RJC72,Ell72} was last listed in 1974 
	\cite{PDG74}. Replacing it by $f_0(500)$ was proposed in 1996 
	\cite{Torn96}.}
but it was already clear that agreement with data required the inclusion 
of large dispersive effects which had to be somehow ``married'' 
to chiral predictions \cite{Don95}. The same can be said now, except that
the $f_0(500)$ pole of Eq.~(\ref{f_0}) can be identified as the source
of these effects. Consequently dispersion theory for these processes, with 
the possible exception of $\eta \to 3\pi$ decay \cite{Col11}, is far better 
understood \cite{Pol02,Ynd07,Pen06,Col12,Trof12}. 

But that does nothing to alter the fact that the lowest order of 
standard chiral $SU(3)_L \times SU(3)_R$ perturbation theory $\chi$PT$_3$ 
fits these data so poorly. The lowest order amplitude ${\cal A}_\mathrm{LO}$ 
is the first term of an asymptotic series 
\begin{equation} 
{\cal A} = \bigl\{{\cal A}_\mathrm{LO} + {\cal A}_\mathrm{NLO} 
         + {\cal A}_\mathrm{NNLO} + \ldots\bigr\}_{\chi\mathrm{PT}_3} 
\label{chiral}\end{equation}
in powers of $O(m_K)$ momenta and quark masses $m_{u,d,s} = O(m_K^2)$ (with
$m_{u,d}/m_s$ held fixed). If the first term is a poor fit, \emph{any}
truncation of the series to make it agree with a dispersive fit to data 
is unsatisfactory \emph{because the series is diverging}.

For example, consider the amplitude for $K_L \to \pi^0\gamma\gamma$ 
(item 2 above). Let the series (\ref{chiral}) be matched to data by 
including dispersive NLO corrections (next to lowest order) and then
truncating:
\begin{equation}
{\cal A}_{K_L \to \pi^0\gamma\gamma} \simeq  \bigl\{{\cal A}_\mathrm{LO} 
         + {\cal A}_\mathrm{NLO}\bigr\}_{\chi\mathrm{PT}_3}  \,.
\end{equation}
The LO prediction for the rate is 1/3 too small, so, depending on the
relative phase of the LO and NLO terms, a fit can be achieved only for
\begin{equation}
\bigl|{\cal A}_\mathrm{NLO}\bigr|_{\chi\mathrm{PT}_3} \gtrsim 
\sqrt{2}\bigl|{\cal A}_\mathrm{LO}\bigr|_{\chi\mathrm{PT}_3}  \,.
\label{Klong}
\end{equation}

How can this be reconciled with the success \cite{Gasser85} of 
$\chi$PT$_3$ elsewhere? Corrections to lowest order $\chi$PT$_3$ should 
be $\sim$ 30\% at most:
\begin{equation}
 \bigl|{\cal A}_\mathrm{NLO}\bigl/{\cal A}_\mathrm{LO}\bigr|_{\chi\mathrm{PT}_3}
 \lesssim 0.3  \mbox{ , acceptable fit.}
 \label{eqn:XPT3_fit}
\end{equation}
A standard response%
\footnote{LCT thanks Professor H.~Leutwyler for a discussion of this 
point.}
is that there are limits to the applicability of an expansion like 
$\chi$PT$_3$, so failures in a few cases are to be expected.

\begin{figure}
\includegraphics[scale=0.53]{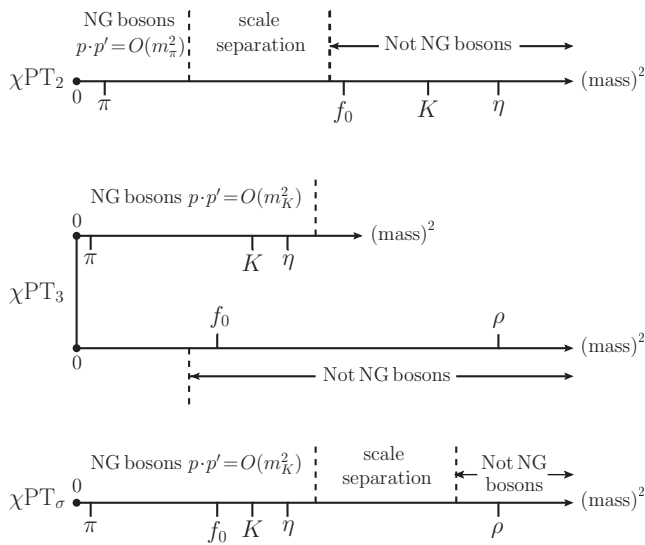}
\caption{Scale separations between Nambu-Goldstone (NG) sectors 
and other hadrons for each type of chiral perturbation theory $\chi$PT 
discussed in this paper. Note that scale separation in $\chi$PT$_2$ 
(chiral $SU(2) \times SU(2)$, top diagram) is ensured 
by limiting extrapolations in momenta $p,p'$ to $O(m_\pi)$ (not $O(m_K)$).
In conventional three-flavor theory  $\chi$PT$_3$ (middle diagram), 
there is {\it no scale separation}: the non-NG boson $f_0(500)$ sits 
in the middle of the NG sector $\{\pi, K, \eta\}$. Our three-flavor
proposal $\chi$PT$_\sigma$ (bottom diagram) for $O(m_K)$ extrapolations
in momenta implies a clear scale separation between the NG sector
$\{\pi,K,\eta,\sigma = f_0\}$ and the non-NG sector $\{\rho,\omega,
K^*\!,N,\eta',\ldots\}$.} 
\label{fig:goldstone}
\end{figure}

In our view, there is a consistent trend of failure in $0^{++}$ channels
which can and should be corrected by modifying the \emph{lowest order} of 
the three-flavor theory.  This must be achieved without changing 
$\chi$PT$_2$, where amplitudes are expanded about the chiral 
$SU(2)_L \times SU(2)_R$ limit with $O(m_\pi)$ extrapolations%
\footnote{For some authors, ``two-flavor theory" refers to pionic processes
\emph{without} the restriction $O(m_\pi)$ on pion momenta. Then the relevant theory is  
$\chi$PT$_3$ or $\chi$PT$_\sigma$, not $\chi$PT$_2$. See 
Fig.~\ref{fig:goldstone}.}
in momenta; $\chi$PT$_2$ is wholly successful, producing convergent results 
with small corrections, typically 5\% or at most 10\%:
\begin{equation}
 \bigl|{\cal A}_\mathrm{NLO}\bigl/{\cal A}_\mathrm{LO}\bigr|_{\chi\mathrm{PT}_2}
 < 0.1 \mbox{ , observed fits.}
\end{equation} 

Our solution is to replace $\chi$PT$_3$ by chiral-scale perturbation theory 
$\chi$PT$_\sigma$, whose NG sector $\{\pi,K,\eta,\sigma/f_0\}$
includes $f_0(500)$ as a dilaton $\sigma$ associated with the scale-invariant
limit (\ref{scale}). In $\chi$PT$_\sigma$, the strange quark mass $m_s$ 
sets the scale of $m^2_{f_0}$ as well as $m^2_K$ and $m^2_\eta$ 
(Fig.~\ref{fig:goldstone}, bottom diagram). As a result, the rules for 
counting powers of $m_K$ are changed:  $f_0$ pole amplitudes (NLO in  
$\chi$PT$_3$) are promoted to LO. That fixes the LO problem for amplitudes 
involving $0^{++}$ channels and $O(m_K)$ extrapolations in momenta. At 
the same time, $\chi$PT$_\sigma$ \emph{preserves} the LO successes of  
$\chi$PT$_3$ elsewhere: for reactions which do not involve $\sigma/f_0$,
the predictions of $\chi$PT$_3$ and $\chi$PT$_\sigma$ are identical.

The analysis relies on a clear distinction being drawn between 
$\chi$PT$_2$, $\chi$PT$_3$, and $\chi$PT$_\sigma$. For each amplitude 
$\cal A$, these three versions of $\chi$PT produce three inequivalent 
asymptotic expansions of the form (\ref{chiral}). The corresponding 
scale separations between NG sectors and other particles
are shown in Fig.~\ref{fig:goldstone}.

We use $\chi$PT$_2$ in the strict sense originally intended
\cite{Wein66,DaWe69,Pagels75,Gasser83,Gasser84}: an asymptotic expansion for 
the limit $m_{u,d} \to 0$ with $m_s \not= 0$ and (crucially) momentum 
extrapolations limited to $O(m_\pi)$. There are only three NG bosons 
$\{\pi^+, \pi^0, \pi^-\}$, with \emph{no dilaton}: $\chi$PT$_2$
is not sensitive to the behavior of $\beta$ because of the relatively 
large term $m_s\bar{s}s$ in Eq.~(\ref{eqn:anomaly}) for $\theta^\mu_\mu$.
Since $s$ is not treated as a light quark, the $K$ and $\eta$ mesons as 
well as $f_0, \rho, \omega, N, \eta' \ldots$ are excluded from the 
$\chi$PT$_2$ NG sector.

If there is an $O(m_K)$ extrapolation in momentum, $\chi$PT$_2$ is 
\emph{not} sufficient. Three-flavor contributions must be included, 
either as large dispersive extrapolations, or with $\chi$PT$_2$ 
replaced by a three-flavor chiral expansion: $\chi$PT$_3$
\cite{Pagels75,GMOR68,Gasser82,Gasser85,Scherer12} or $\chi$PT$_\sigma$. 

An $O(m_K)$ extrapolation may arise because $K$ 
or $\eta$ is soft, or because the pion momenta in (say) $\pi\pi \to 
\pi\pi$ or $\gamma\gamma \to \pi\pi$ are chosen to be $O(m_K)$, or 
because of a kinematic constraint. A well known example is the fact 
that  $\chi$PT$_2$ says almost nothing about $K_S \to \pi\pi$: if one pion 
becomes soft, the momentum difference between on-shell states 
$|K\rangle$ and $|\pi\rangle$ is necessarily $O(m_K)$. An example of 
interest in Sec.~\ref{weak_emag} is the pion-loop result \cite{DAm86} 
for $K_S \to \gamma\gamma$, which is not implied by $\chi$PT$_2$:
a three-flavor expansion is necessary.

Both $\chi$PT$_3$ and $\chi$PT$_\sigma$ involve the limit%
\footnote{We require $m_s > m_{u,d}$ throughout.
Double asymptotic series can be considered for either $\chi$PT$_2$ and 
$\chi$PT$_3$ \cite{Gasser85,Gasser07} or $\chi$PT$_2$ and $\chi$PT$_\sigma$.  
The unusual limit $m_s \to 0$ for fixed $m_{u,d} \not= 0$ considered in 
Sec.~4 of \cite{Nebreda10} does not produce any NG bosons.}
\begin{equation}
 m_i \sim 0\ ,\ m_i/m_j \mbox{ fixed, } i,j = u,d,s.
\label{3-flavor}\end{equation}
In each case, amplitudes are expanded in powers and logarithms of 
\begin{equation}
 \{\mbox{momenta}\}\bigl/\chi_\mathrm{ch} \ll 1 
\label{irscale}\end{equation}
where the infrared mass scale $\chi_\mathrm{ch} \approx 1 \mbox{ GeV}$ is 
set by the chiral condensate $\langle \bar{q}q \rangle_{\mathrm{vac}}$.
In $\chi$PT$_3$,  $\chi_\mathrm{ch}$ is $4\pi F_\pi$ \cite{ManGeo}, 
where $F_\pi = 93$ MeV is the pion decay constant;
a similar result will be found for $\chi$PT$_\sigma$ in Sec.~\ref{strong}.
The chiral scale  $\chi_\mathrm{ch}$ also sets the mass scale of particles 
outside the corresponding NG sectors.%
\footnote{Except for glueballs, if they exist. In $\chi$PT$_\sigma$, they 
may have large masses due to gluonic scale condensates such as 
$\langle G^2 \rangle_\mathrm{vac}$.}
For nucleons with mass $M_N$, this is evident from the Goldberger-Treiman 
relation
\begin{equation}
F_\pi g_{\pi NN} \simeq g_A M_N  \,.
\label{GT}\end{equation}

It is essential \cite{ManGeo} to make a clear distinction between 
the low-energy scale $\chi_\mathrm{ch}$ and the ultraviolet QCD scale 
$\Lambda_\mathrm{QCD} \approx 200$ MeV associated with expansions 
in the asymptotically free domain
\begin{equation}
 \{\mbox{momenta}\}\bigl/\Lambda_\mathrm{QCD}  \gg 1. 
\label{uvscale}\end{equation}
Strong gluonic fields are presumably responsible for both scales, but 
that does not mean that the dimensionless ratio
\begin{equation}
\chi_\mathrm{ch}\bigl/\Lambda_\mathrm{QCD} \approx 5
\end{equation}
has to be 1. 

The difference between $\chi$PT$_3$ and $\chi$PT$_\sigma$ 
can be seen in the relation between hadronic masses and terms in 
Eq.~(\ref{eqn:anomaly}) for $\theta^\mu_\mu$.

In $\chi$PT$_3$, there is no sense in which the gluonic trace 
anomaly is small. For example, the gluonic anomaly is taken
to be responsible for most of the nucleon's mass:
\begin{equation}
M_N = \langle N|\theta^\mu_\mu |N\rangle
    \underset{\chi\mathrm{PT}_3}{=} 
\frac{\beta(\alpha_s)}{4\alpha_s}\langle N|G^2 |N\rangle 
       + O\bigl(m_K^2\bigr) \,. 
\end{equation}
This assumes that $f_0(500)$ pole terms can be neglected, or equivalently,
given that $f_0$ is so light on the mass scale 
for non-NG particles set by $\chi_\mathrm{ch}$, that $f_0$ couples weakly to 
$G^2$ and $\bar{q}q$. As noted in Fig.~\ref{fig:goldstone}, the small
$f_0$ mass implies that $\chi$PT$_3$ has no scale separation, which 
(as we have seen) is a problem because $f_0$ couples so strongly 
to other particles.

Contrast this with $\chi$PT$_\sigma$, where the infrared regime
\begin{equation}
O(m_K) \mbox{ momenta } \ll \chi_\mathrm{ch} 
\end{equation}
emphasizes values of $\alpha_s$ close to $\alpha^{}_\mathrm{IR}$, so a
combined limit
\begin{equation}
m_{u,d,s} \sim 0 \quad\mbox{and}\quad \alpha_s \lesssim \alpha^{}_\mathrm{IR}  
\end{equation}
must be considered. Since $\beta(\alpha_s)$ is small, the gluonic trace 
anomaly is small \emph{as an operator}, but it can produce large amplitudes 
when coupled to dilatons.

\begin{figure}[t]
\center\includegraphics[scale=.7]{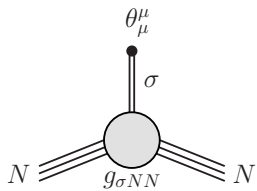}
\caption{Dominant $\sigma$-pole diagram in $\chi$PT$_\sigma$ for 
$\langle N|\theta^\mu_\mu |N\rangle$.} 
\label{fig:g_sigNN}
\end{figure}%

Consider how $M_N$ arises in $\chi$PT$_\sigma$ 
(Fig.~\ref{fig:g_sigNN}). Like other pseudo-NG bosons, $\sigma$ 
couples to the vacuum via the divergence of its symmetry current,
\begin{equation}
 \langle\sigma | \theta^\mu_\mu |\mbox{vac}\rangle = - m_\sigma^2 F_\sigma
= O(m_\sigma^2)\,, \ m_\sigma \to 0
\label{dilaton}\end{equation}
where $F_\sigma$ is the dilaton decay constant. The nucleon remains massive 
in the scaling limit because of its coupling $-g_{\sigma NN}\sigma\bar{N}N$ 
to $\sigma$ and the factor $-i/m_\sigma^2$ produced by the $\sigma$ pole
at zero momentum transfer. This gives rise to the well known analogue
\cite{Carr71}
\begin{equation}
 F_\sigma g_{\sigma NN} \simeq M_N
\label{scalarGT}\end{equation}
of the Goldberger-Trieman relation (\ref{GT}).

In our scheme, both the gluonic anomaly and the quark mass term in 
Eq.~(\ref{eqn:anomaly}) for $\theta^\mu_\mu$ can contribute to $M_N$ in 
the chiral-scale limit (\ref{scale}). That is because we require%
\footnote{In principle, we could have constructed a chiral-scale 
perturbation theory with $m_\sigma$ and $m_K$ as independent expansion 
parameters, but that would make sense only if there were a fourth light 
quark or different low-energy scales for chiral and scale expansions. 
Fig.~\ref{fig:goldstone} provides clear confirmation that the choice 
$m_\sigma = O(m_K)$ is sensible.\label{one_scale}}
\begin{equation}
m_\sigma^2 = O(m_K^2) = O(m_{u,d,s})\,, 
\label{dilaton-mass}
\end{equation}
which allows the constants $F_{G^2}$ and $F_{\bar{q}q}$ given by
\begin{align}
\beta(\alpha_s)\bigl/(4\alpha_s)\langle\sigma | G^2 |\mbox{vac}\rangle 
  &=  - m_\sigma^2 F_{G^2}  \,,  \nonumber \\
      \{1 + \gamma_m(\alpha_s)\}\sum_{q = u,d,s}m_q
         \langle\sigma | \bar{q}q |\mbox{vac}\rangle 
  &=  - m_\sigma^2 F_{\bar{q}q}  
\label{constants}\end{align}
to remain finite in that limit:
\begin{equation}
M_N \simeq F_{G^2}g_{\sigma NN} + F_{\bar{q}q}g_{\sigma NN} \,.
\end{equation}

Suggestions that a resonance like $f_0(500)$ cannot be a pseudo-NG boson
have no foundation. There can be no theorem to that effect because 
counterexamples such as our effective chiral-scale Lagrangian in 
Sec.~\ref{Lagrangian} are so easily constructed. It is true that in the 
symmetry limit where a NG boson becomes exactly massless, it has zero width, 
but that is because there is no phase space for it to decay into other
massless particles. If phase space for strong decay is made available by
explicit symmetry breaking and quantum number conservation allows it, a 
pseudo-NG boson will decay:
\begin{equation} 
m_\sigma > 2 m_\pi \ \Rightarrow\  \mbox{ width } 
 \Gamma_{\sigma \to \pi\pi} \not= 0 \,.
 \label{eqn:phase}
\end{equation}
Note that:
\begin{itemize}
\item Non-NG bosons need not be resonances; for example, $\eta'(960)$ is
stable against strong decay.
\item The resonance $f_0/\sigma$ becomes a massless NG boson \emph{only} if 
all three quarks $u,d,s$ become massless as $\alpha_s$ tends to 
$\alpha^{}_\mathrm{IR}$. In that combined limit, all particles except 
$\pi,K,\eta$ and $\sigma$ remain massive. Strong gluon fields set the 
scale of the condensate $\langle \bar{q}q \rangle_{\mathrm{vac}}$, which 
then sets the scale for massive particles and resonances except (possibly) 
glueballs.
\item QCD at $\alpha_s = \alpha^{}_\mathrm{IR}$ resembles the physical theory 
(i.e.\ QCD for $0 < \alpha_s < \alpha^{}_\mathrm{IR}$) in the resonance region, 
but differs completely at high energies because it lacks asymptotic freedom. 
Instead, Green's functions scale asymptotically with nonperturbative 
anomalous dimensions in the ultraviolet limit.
\end{itemize}

Another key difference between $\chi$PT$_3$ and  $\chi$PT$_\sigma$ becomes
evident in the many-color limit $N_c \to \infty$ 
\cite{tHooft,Venez,Witten}. At issue is the quark content of the $f_0(500)$ 
resonance: is it a standard $q\bar{q}$ meson, or an exotic tetraquark state 
$q\bar{q}q\bar{q}$?  
In general, this is a model-dependent question; indeed the tetraquark idea 
was first proposed for the $0^+$ nonet in the context of the quark-bag model 
\cite{Jaffe}. However the large-$N_c$ limit permits conclusions which are
far less model-dependent.

In modern analyses of $\chi$PT$_3$, $f_0(500)$ is often considered 
to be a multi-particle state and so is not represented by a field in 
an effective Lagrangian. Instead, the $\chi$PT$_3$ expansion is unitarized, 
with $f_0$ identified as a resonating two-meson state produced by 
the unitarized structure. From that, the large-$N_c$ conclusion 
\cite{Pelaez11} 
\begin{equation}
f_0 \sim \pi\pi \sim (q\bar{q})^2 \,, \ \mbox{ unitarized $\chi$PT$_3$}
\end{equation}
is drawn. This assumes from the outset that $f_0$ is \emph{not} a 
dilaton. The problem, already discussed at the beginning of this Section,
is that the $\chi$PT$_3$ expansion diverges because it is dominated by these 
unitary ``corrections''\!. 

In $\chi$PT$_\sigma$, the large-$N_c$ properties of $f_0/\sigma$ are
similar to those of pions, and are found by considering the two-point function
of $\theta_{\mu\nu}$ instead of chiral currents. At large-$N_c$, the spin-2
part is dominated by pure-glue states:
\begin{equation}
T\bigl\langle\mathrm{vac}\bigl|\theta_{\alpha\beta}\theta_{\mu\nu}
     \bigr|\mathrm{vac}\bigr\rangle^{}_\mathrm{spin-2} = O(N_c^2) \,.
\end{equation}
However, when the spin-0 part is projected out by taking the trace
$\theta_\alpha^\alpha$, the quark term dominates the gluonic anomaly
of Eq.~(\ref{eqn:anomaly}) at large $N_c$ because of the factor 
$\alpha_s \sim 1/N_c$ multiplying $G^2$. Thus we find 
\begin{equation}
T\bigl\langle\mathrm{vac}\bigl|\theta^\alpha_\alpha\theta^\mu_\mu
     \bigr|\mathrm{vac}\bigr\rangle = O(N_c) 
\label{2pt}\end{equation}
due to the quark term compared with $O(1)$ from the gluonic anomaly. 
Clearly, a $\sigma$ pole can be present only if $f_0/\sigma$ is a 
$q\bar{q}$ state. At zero momentum transfer, this pole contributes
$m_\sigma^2F_\sigma^2$ to the amplitude (\ref{2pt}), from which we conclude
\begin{equation}
F_\sigma = O\bigl(\sqrt{N_c}\bigr) \,,
\label{Fsigma}
\end{equation}
as for the pion decay constant $F_\pi$. We will see in Sec.~\ref{strong} 
that the dilaton, like other $q\bar{q}$ states, obeys the narrow width rule 
at large $N_c$.

Sometimes pure-glue corrections in $f_0/\sigma$ are dominant. The most obvious
example is the nucleon mass $M_N$, where the leading $O(N_c)$ contribution
due to $q\bar{q}$ states is the numerically small two-flavor sigma term
\begin{equation}
\langle N|m_u\bar{u}u + m_d\bar{d}d|N \rangle \ll M_N \,.
\end{equation}
Therefore (as is generally agreed) most of $M_N$ comes from the 
$m_{u,d}$-independent term due to pure-glue exchange. In particular, 
the terms $\sim G^2$ and $m_s\bar{s}s$ in Eq.~(\ref{eqn:anomaly}) for 
$\theta^\mu_\mu$ couple to a nucleon only via pure-glue states.

\section{Chiral-scale Lagrangian}
\label{Lagrangian}
Consider strong interactions at low energies 
$\alpha_s \lesssim \alpha^{}_\mathrm{IR}$ within the physical region
\begin{equation} 
0 < \alpha_s < \alpha^{}_{\mathrm{IR}} \,.
\label{phys}\end{equation}
Let $d$ denote the scaling dimension of operators used to construct
an effective chiral-scale Lagrangian. In general, there must be a
scale-invariant term $\mathcal{L}_\mathrm{inv}$ with scaling dimension 
$d = 4$, a term $\mathcal{L}_\mathrm{mass}$ with dimension \cite{Wilson69} 
\begin{equation} 
d_\mathrm{mass} = 3 - \gamma_{m}\bigl(\alpha^{}_\mathrm{IR}\bigr)\ , \quad
1 \leqslant d_\mathrm{mass} < 4
\label{dim-mass}
\end{equation} 
to simulate explicit breaking of chiral symmetry by the quark mass term, 
and a term $\mathcal{L}_\mathrm{anom}$ with dimension $d > 4$ to account 
for gluonic interactions responsible for the strong trace anomaly in 
Eq.~(\ref{eqn:anomaly}):
\begin{equation} 
\mathcal{L}^{}_{\mbox{\small $\chi$PT$_\sigma$}}
 =\ :\mathcal{L}^{d=4}_\mathrm{inv} 
 + \mathcal{L}^{d>4}_\mathrm{anom} + \mathcal{L}^{d<4}_\mathrm{mass}: \,.
\label{Lagr}\end{equation}
The anomalous part of $d_\mathrm{mass}$ is evaluated at $\alpha^{}_\mathrm{IR}$ 
because we expand in $\alpha_s$ about $\alpha^{}_\mathrm{IR}$. A proof 
that $\mathcal{L}_\mathrm{anom}$ has dimension $d > 4$ appears later 
in this Section.

We restrict our analysis to the NG sector of $\chi$PT$_\sigma$ 
(Fig.~\ref{fig:goldstone}). Then operators in
\begin{equation} 
\mathcal{L}^{}_{\mbox{\small $\chi$PT$_\sigma$}}
= \mathcal{L}\bigl[\sigma,U,U^\dagger\bigr]
\end{equation}
are constructed from
a QCD dilaton field $\sigma$ and the usual chiral $SU(3)$ field
\begin{equation}
U = U(\pi,K,\eta)\,, \ UU^\dagger = I  \,.
\end{equation}
Scale and chiral transformations commute, so $\sigma$ is chiral invariant.
The scale dimensions of $\pi,K,\eta$ and hence $U$ must be zero in order 
to preserve the range of field values on the coset space 
$SU(3)_L\times SU(3)_R/SU(3)_V$ \cite{Chiv89}.
 
In Eq.~(\ref{Lagr}), both $\mathcal{L}_\mathrm{inv}$ and 
$\mathcal{L}_\mathrm{anom}$ are $SU(3)_L \times SU(3)_R$ invariant, 
while $\mathcal{L}_\mathrm{mass}$ belongs to the representation 
$(\mathbf{3},\bar{\mathbf{3}})\oplus(\bar{\mathbf{3}},\mathbf{3})$ 
associated with the $\pi,K,\eta$ (mass)$^2$ matrix
$M$. In lowest order, with $M$ diagonalized, 
\begin{equation}
M = \frac{F_\pi^2}{4}\left(\begin{matrix} 
        m_\pi^2 & 0 & 0 \\ 0 & m_\pi^2 & 0 \\ 0 & 0 & 2m_K^2 - m_\pi^2 
                      \end{matrix}\right) 
\label{mass_matrix}\end{equation}
the vacuum condition for $U$ is
 \begin{equation}
U \to I \ \mbox{ for } \ \pi,K,\eta \to 0 \,.
\label{vac_cond}
\end{equation}

The dimension of $\mathcal{L}_\mathrm{anom}$ can be found
from the scaling Ward identities (Callan-Symanzik equations)
\begin{equation}
\bigg\{ \mu\frac{\del}{\del\mu} + \beta(\alpha_s)\frac{\del}{\del\alpha_s} + 
\gamma_{m}(\alpha_s)\sum_q m_q\frac{\del\ }{\del m_q} \bigg\}{\cal A} = 0
\label{CSeqn}
\end{equation}
for renormalization-group invariant QCD amplitudes $\mathcal{A}$. The term 
$\beta\del/\del\alpha_s$ corresponds to the gluonic anomaly in 
Eq.~(\ref{eqn:anomaly}), so the effect of $\alpha_s\del/\del\alpha_s$
on $\mathcal{A}$ is to insert the QCD operator $G^2 = G_{\mu\nu}^aG^{a\mu\nu}$
at zero momentum transfer. Applying $\alpha_s\del/\del\alpha_s$
to Eq.~(\ref{CSeqn}), 
\begin{align}
\biggl\{ \mu\frac{\del\ }{\del\mu} + &\beta(\alpha_s)\frac{\del\ }{\del\alpha_s}
+ \beta'(\alpha_s) - \beta(\alpha_s)\bigl/\alpha_s\biggr\}
\alpha_s\frac{\del{\cal A}}{\del\alpha_s} \nonumber \\
&=  -\alpha_s\frac{\del\ }{\del\alpha_s} \sum_q \gamma_{m}(\alpha_s)
    m_q\frac{\del{\cal A_{}}}{\del m_q}
\label{del_A}
\end{align}
we see that the anomalous dimension function for $G^2$ is 
\begin{equation}
\gamma^{}_{G^2}(\alpha_s) = \beta'(\alpha_s) - \beta(\alpha_s)/\alpha_s \,.
\label{gamma}
\end{equation}
Hence, to lowest order in the expansion $\alpha_s \lesssim 
\alpha^{}_\mathrm{IR}$, ${\cal L}_\mathrm{anom}$ has a positive anomalous 
dimension equal to the slope of $\beta$ at the fixed point 
(Fig.~\ref{fig:beta}):
\begin{equation}
d_\mathrm{anom} = 4 + \beta'\bigl(\alpha^{}_\mathrm{IR}\bigr) > 4\,.
\label{dim-anom}\end{equation}

As $\alpha_s \to \alpha^{}_\mathrm{IR}$, the gluonic anomaly vanishes, 
so for consistency,\footnotemark[10] we 
must require terms in ${\cal L}_\mathrm{anom}$ to involve derivatives 
$\del\del = O(M)$ or have $O(M)$ coefficients:
\begin{equation}
{\cal L}_\mathrm{anom} = O(\del^2, M) \,.
\label{L_anom}\end{equation}
The result is a chiral-scale perturbation expansion $\chi$PT$_\sigma$ 
about $\alpha^{}_\mathrm{IR}$ with QCD dilaton mass $m_\sigma = O(m_K)$.

An explicit formula for the $\chi$PT$_\sigma$ Lagrangian (\ref{Lagr}) can
be readily found by following the approach of Ellis \cite{Ell70,Ell71}. 
Let $F_\sigma$ be the coupling of $\sigma$ to the vacuum via the energy 
momentum tensor $\theta_{\mu\nu}$, improved \cite{CCJ70} when spin-0 fields 
are present:
\begin{equation} 
\langle\sigma(q)|\theta_{\mu\nu}|\mathrm{vac}\rangle
= (F_\sigma/3)\bigl( q_\mu q_\nu - g_{\mu\nu}q^2\bigr) \,.
\label{b}\end{equation}
When conformal symmetry is realized nonlinearly \cite{Salam}, a dilaton 
field $\sigma$ is needed to create connection terms $\sim \del\sigma$ in 
covariant derivatives. It transforms as
\begin{equation}
\sigma \to \sigma - \tfrac{1}{4}F_\sigma \log \big|\det (\del x'/\del x) \big|
\label{sigma_scale}\end{equation}
under conformal transformations $x \to x'$, which corresponds to scale
dimension 1 for the covariant field $e^{\sigma/F_\sigma}$. The dimensions of 
$\chi$PT$_3$ Lagrangian operators such as 
\begin{equation}
\mathcal{K}\bigl[U,U^\dagger\bigr] 
= \tfrac{1}{4}F_{\pi}^{2}\mathrm{Tr}(\partial_{\mu} U\partial^{\mu}U^{\dagger})
\end{equation}
and the dilaton operator 
$\mathcal{K}_\sigma = \frac{1}{2}\partial_{\mu}\sigma\partial^{\mu}\sigma$
can then be adjusted by powers 
of $e^{\sigma/F_\sigma}$ to form terms in $\mathcal{L}$. In lowest order,
\begin{align}
&\mathcal{L}^{d=4}_\mathrm{inv,\,LO}
 = \bigl\{c_{1}\mathcal{K} + c_{2}\mathcal{K}_\sigma 
     + c_{3}e^{2\sigma/F_{\sigma}}\bigr\}e^{2\sigma/F_{\sigma}} \,,
\notag \\[1mm]
&\mathcal{L}^{d>4}_\mathrm{anom,\,LO} \notag \\
&= \bigl\{(1-c_{1})\mathcal{K} + (1-c_{2})\mathcal{K}_\sigma
      + c_4 e^{2\sigma/F_{\sigma}}\bigr\}e^{(2+\beta')\sigma/F_{\sigma}} \,,
\notag \\[1mm]
&\mathcal{L}^{d<4}_\mathrm{mass,\,LO} 
 = \mathrm{Tr}(MU^{\dagger}+UM^{\dagger})e^{(3-\gamma_{m})\sigma/F_{\sigma}} \,,
\label{Lstr}
\end{align}
where $\beta'$ and $\gamma_{m}$ are the anomalous dimensions 
$\beta'(\alpha^{}_\mathrm{IR})$ and $\gamma_{m}(\alpha^{}_\mathrm{IR})$ 
of Eqs.\ (\ref{dim-anom}) and (\ref{dim-mass}).  

The constants $c_{1}$ and $c_{2}$ are not fixed by general 
arguments, while $c_3$ and $c_4$ depend on the vacuum condition 
chosen for the field $\sigma$. The role of $c_3$ and $c_4$ is to
fix the scale of $e^{\sigma/F_\sigma}$, just as the (mass)$^2$ matrix 
fixes the chiral $SU(3)$ direction of $U$ (Eqs.~(\ref{mass_matrix}) 
and (\ref{vac_cond})). The simplest choice of field variables%
\footnote{On-shell amplitudes do not depend on how 
the field variables are chosen \cite{Chi61,Kam61}.}
is to have all NG fields $\sigma, \pi, K, \eta$ fluctuate about 
zero.

For the vacuum to be stable in the $\sigma$ direction at $\sigma = 0$, 
Lagrangian terms linear in $\sigma$ must cancel:
\begin{align}
4c_3 + (4+\beta')c_4 &= - (3-\gamma_{m})\bigl\langle\mathrm{Tr}
        (MU^{\dagger}+UM^{\dagger})\bigr\rangle_{\mathrm{vac}} \notag \\
 &= - (3-\gamma_{m})F_\pi^2\bigl(m_K^2 + \tfrac{1}{2}m_\pi^2\bigr)\,.
\label{stable}\end{align}
Eqs.~(\ref{L_anom}) and (\ref{stable}) imply that both $c_3$ and 
$c_4$ are $O(M)$. 

Evidently $\chi$PT$_\sigma$ is a simple extension of the conventional
three-flavor theory $\chi$PT$_3$. The  $\chi$PT$_\sigma$ Lagrangian defined 
by Eqs.~(\ref{Lagr}) and (\ref{Lstr}) satisfies the condition
\begin{equation}
{\cal L}^{}_{\mbox{\small $\chi$PT$_\sigma$}} 
 \to {\cal L}^{}_{\mbox{\small $\chi$PT$_3$}} \,, \
\sigma \to 0 
\end{equation}
and hence preserves the phenomenological success of \emph{lowest order} 
$\chi$PT$_3$ for amplitudes which do not involve the $0^{++}$ channel 
(Sec.~\ref{Motiv}).  In next to lowest order, new chiral-scale loop 
diagrams involving $\sigma$ need to be checked.

The $\chi$PT$_\sigma$ Lagrangian obeys the standard rule that each
term ${\cal L}_d$ of dimension $d$ contributes $(d-4){\cal L}_d$ to the 
trace of the effective energy-momentum tensor:
\begin{equation}
\left.\theta^\mu_\mu\right|_\mathrm{eff} 
   =\ :\beta'\mathcal{L}^{d>4}_\mathrm{anom} 
      - (1+\gamma_{m})\mathcal{L}^{d<4}_\mathrm{mass}: \,. \label{eff-tr}
\end{equation}
Note that the critical exponent $\beta'$ normalizes the gluonic term in
$\theta^\mu_\mu$.

\section{Strong Interactions}
\label{strong}
In lowest order, $\cal L$ gives formulas for the $\sigma\pi\pi$ coupling 
\begin{equation}
\mathcal{L}_{\sigma\pi\pi}
 = \bigl\{\bigl(2+(1-c_1)\beta'\bigr)|\del\bm{\pi}|^2 
    - (3 - \gamma_{m})m_\pi^2|\bm{\pi}|^2\bigr\}\sigma/(2F_\sigma) \label{Lsigpi}
\end{equation}
and dilaton mass $m_\sigma$
\begin{equation}
m_\sigma^2 F_\sigma^2
= F_\pi^2\bigl(m_K^2 + \tfrac{1}{2}m_\pi^2\bigr)(3 - \gamma_{m})(1 + \gamma_{m})
  - \beta'(4 + \beta')c_4
\label{mass}\end{equation}
which resemble pre-QCD results \cite{Ell70,RJC70,Ell71,Klein71} but have
extra gluonic terms proportional to $\beta'$. For consistency with data, 
we must assume that the unknown coefficient $2+(1-c_1)\beta'$ in 
Eq.~(\ref{Lsigpi}) does not vanish accidentally. That preserves the 
key feature of the original work, that $\mathcal{L}_{\sigma\pi\pi}$ is 
mostly \emph{derivative}: for soft $\pi\pi$ scattering (energies 
$\sim m_\pi$), the dilaton pole amplitude is negligible because the 
$\sigma\pi\pi$ vertex is $O(m_\pi^2)$, while the  $\sigma\pi\pi$ vertex 
for an on-shell dilaton
\begin{equation}
g_{\sigma\pi\pi} = -\bigl(2+(1-c_1)\beta'\bigr)m_\sigma^2/(2F_\sigma) 
+ O(m_\pi^2)
\label{on-shell}
\end{equation}
is $O(m_\sigma^2)$, consistent with $\sigma$ being the broad resonance 
$f_0(500)$.

Comparisons with data require an estimate of $F_\sigma$, most simply from
$NN$ scattering and the dilaton relation (\ref{scalarGT}).
The data imply \cite{CC08} a mean value $g_{\sigma NN} \sim 9$ and 
hence $F_{\sigma} \sim 100\,\mathrm{MeV}$ but with an uncertainty
which is either model-dependent or very large ($\approx 70\%$).  That 
accounts for the large uncertainty in 
\begin{equation}
 1\tfrac{1}{2} \lessapprox |2 + (1-c_{1})\beta'| \lessapprox 6
\label{inequal}
\end{equation}
when we compare Eq.~(\ref{on-shell}) with data \cite{Cap06}: 
\begin{equation}
|g_{\sigma\pi\pi}| = 3.31^{+0.35}_{-0.15}\mbox{ GeV, and }
m_{\sigma} \approx 441\,\mathrm{MeV}. 
\label{numbers}
\end{equation}

The convergence of a chiral-scale expansion can be tested by adding 
$\sigma$-loop diagrams to the standard  $\chi$PT$_3$ analysis 
\cite{Gasser85}. These involve the (as yet) undetermined constants
$\beta',\gamma_{m},c_{1\ldots 4}$: for example, corrections to $g_{\sigma\pi\pi}$
involve the $\sigma\sigma\sigma$ and $\sigma\sigma\pi\pi$ vertices derived
from Eq.~(\ref{Lstr}). 

\begin{figure}[t]
\center\includegraphics[scale=0.6]{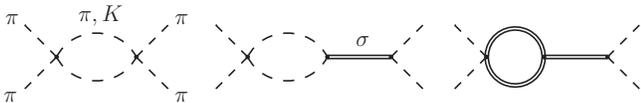}
\caption{Examples of NLO $\chi$PT$_\sigma$ graphs in the chiral-scale 
expansion of $\pi\pi$ scattering for $O(m_K)$ momenta. Each vertex is 
generated by the lowest order terms (\ref{Lstr}) in $\mathcal{L}$.  
Not shown are additional diagrams involving the self-energy of the 
$\sigma$ propagator, and internal $\sigma$ lines which connect one 
external $\pi$ leg to another. Similar diagrams are found for the 
$t$ and $u$ channels.}
\label{fig:pipi_scatter}
\end{figure}

However a numerical estimate of scales associated with the expansion 
can be obtained using the dimensional arguments of Manohar and Georgi 
\cite{ManGeo}. The idea is to count powers of dimensionful quantities 
$F_\pi$ and (for $\chi$PT$_\sigma$) $F_\sigma$ associated with the quark 
condensate $\langle \bar{q}q \rangle_{\mathrm{vac}}$, and keep track of 
powers of $4\pi$ arising from loop integrals. To illustrate their 
point, Manohar and Georgi considered loop corrections to $\pi\pi$ 
scattering, such as the first diagram in Fig.\ \ref{fig:pipi_scatter}, 
for which they obtained the estimate  
\begin{equation}
{\cal A}_\mathrm{loop}\bigl/{\cal A}_\mathrm{tree} 
 \sim \frac{1}{16\pi^2 F_\pi^2} \times \mbox{logarithms}.
\end{equation}
In our scheme, we must add contributions
\begin{equation}
\sim \biggl\{\frac{1}{16\pi^2 F^2_\sigma}\,\mbox{ and }
   \frac{F_\pi^2}{16\pi^2 F_\sigma^4}\biggr\} \times \mbox{logarithms}
\end{equation}
from e.g.\ the second and third graphs of Fig.\ \ref{fig:pipi_scatter}.
As a result, we find that there are in principle \emph{two} $\chi$PT$_\sigma$ 
scales
\begin{equation}
\chi_\pi = 4\pi F_\pi \ \mbox{ and } \ \chi_\sigma = 4\pi F_\sigma\,.
\end{equation}
The rough estimate of 100 MeV for $F_\sigma$ (close to $F_\pi \simeq$
93 MeV) indicates that in effect, there is a single infrared mass scale
\begin{equation}
\chi_\pi \approx \chi_\sigma \approx 1 \mbox{ GeV}
\end{equation}
as foreshadowed in Eq.~(\ref{irscale}).

Numerology which ignores factors of $4\pi$ can be as misleading in 
$\chi$PT$_\sigma$ as in $\chi$PT$_3$. The most important example of this 
arises from the observation that $f_0(500)$ is almost as broad as it is heavy. 
Does this mean that the width of $f_0(500)$ is a \emph{lowest} order 
effect, i.e.\ of the same order in $m_\sigma$ as the real part of the mass? 
If so, would not that invalidate PCDC (partial conservation of the 
dilatation current), where dominance by a \emph{real} pole is assumed for the 
lowest order?

To see that the answer is ``no'', let us estimate the $\sigma$ width 
$\Gamma_{\sigma\pi\pi}$ in the spirit of Manohar and Georgi. We find
\begin{equation}
      \Gamma_{\sigma\pi\pi} \approx \frac{|g_{\sigma\pi\pi}|^2}{16\pi m_\sigma} 
      \sim \frac{m_\sigma^3}{16\pi F_\sigma^2} \sim  250 \mbox{ MeV} 
\label{width}\end{equation}
so $\Gamma_{\sigma\pi\pi}$ is $O(m_\sigma^3)$ and hence \emph{nonleading}
relative to the mass $m_\sigma$. We are therefore justified in using
just tree diagrams to generate the lowest order%
\footnote{Beyond lowest order, and in degenerate cases like the 
       $K_{L}$--$K_{S}$ mass difference, methods used to estimate 
       corrections at the $Z^0$ peak \cite{ZOphys} and the $\rho$ 
       resonance \cite{Scherer} may be necessary.\label{degen}}
of $\chi$PT$_\sigma$, as in $\chi$PT$_2$ and $\chi$PT$_3$. (The main 
exception to this rule, for two-photon channels, is discussed in 
Sec.~\ref{Electromag} and Appendix \ref{AppB}.)  Pure numerology fails 
because $F_\sigma$ in the denominator of (\ref{width}) is 
an order of magnitude smaller than $\chi_{\pi,\sigma}$.  

In the large-$N_c$ limit, as shown in Sec.~\ref{Motiv},
the dilaton behaves as a $q\bar{q}$ state. It follows that the gluonic 
corrections $\sim (1-c_1)\beta'$ in  Eq.~(\ref{on-shell}) for the 
$\sigma\pi\pi$ coupling correspond to disconnected quark diagrams,
so they are nonleading
\begin{equation}
(1-c_1)\beta' = O\bigl(1\bigl/N_c\bigr)
\end{equation}
and the pre-QCD result \cite{Ell70,RJC70} 
\begin{equation}
F_\sigma g_{\sigma\pi\pi} \approx - m_\sigma^2
\end{equation}
is recovered for $N_c$ large. It follows from Eq.~(\ref{Fsigma})
that $\sigma$ decouples from $\pi\pi$ at large $N_c$:
\begin{equation}
g_{\sigma\pi\pi} = O\bigl(1/\sqrt{N_c}\bigr)  \,.
\end{equation}
Hence, like other $q\bar{q}$ states, the dilaton $\sigma$ obeys
the narrow width rule
\begin{equation}
\Gamma_{\sigma\pi\pi} = O(1/N_c) \,.
\end{equation}

The technique used to obtain Eq.~(\ref{Lstr}) from  $\chi$PT$_3$ works
equally well for higher order terms in strong interactions, and also
for external operators induced by electromagnetic or weak interactions 
(Sects.~\ref{Electromag} and \ref{weak_emag}).

In general, NLO terms in the strong interaction Lagrangian 
$\cal L$ are  $O(\del^4, M\del^2, M^2)$. For example, let us construct 
$O(\del^4)$ terms from the $\chi$PT$_3$ operator 
$(\mbox{Tr}\del U\del U^\dagger)^2$. It has dimension 4 already, so it 
appears unchanged in the scale-symmetric term
\begin{equation}
{\cal L}^{d=4}_{\mathrm{inv,\,NLO}} = \{\mbox{coefficient}\}
(\mbox{Tr}\del U\del U^\dagger)^2 + \ldots
\label{invNLO}\end{equation}
i.e.\ without $\sigma$ field
dependence.  The anomalous term has dimension greater than 4, so it depends 
on $\sigma$:
\begin{equation}
{\cal L}^{d>4}_{\mathrm{anom,\,NLO}} =
 \{\mbox{coefficient}\}\bigl(\mbox{Tr}\del U\del U^\dagger\bigr)^2
                e^{\beta'\sigma/F_\sigma}.
\label{d>4NLO}\end{equation}

The difference between $\chi$PT$_3$ and $\chi$PT$_\sigma$ is summarized
in Fig.~\ref{fig:compare}. See Appendix \ref{AppA} for a discussion of 
\begin{figure}[b]
\label{fig:compare}
\centering\includegraphics[scale=1]{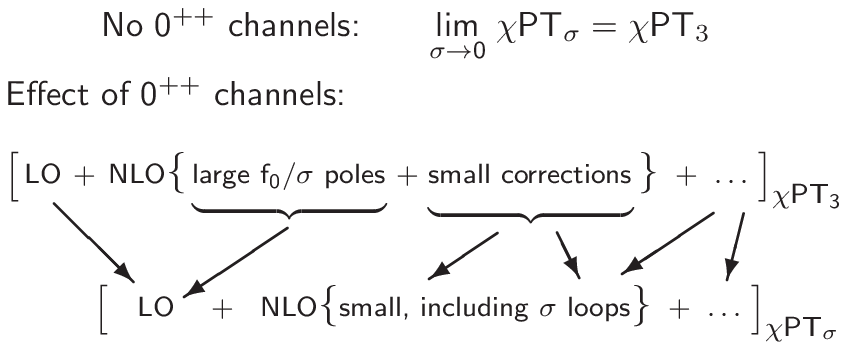}
\caption{Comparison of $\chi$PT$_3$ and $\chi$PT$_\sigma$.
The $f_0/\sigma$ pole terms responsible for the poor convergence of 
$\chi$PT$_3$ are transferred to LO in $\chi$PT$_\sigma$, where they do 
not upset convergence.}
\end{figure}
power counting for $\chi$PT$_\sigma$ loop expansions.

\section{Resonance Saturation in $\bm{\chi}$PT$_\sigma$}
\label{saturate}
Conventional $\chi$PT$_3$ is often supplemented by a technique 
\cite{Eck89} in which the coefficients of $O(\del^4) = O(m_K^4)$ terms are 
estimated by saturation with particles or resonances from the non-NG
sector. This scheme can be readily adapted to $\chi$PT$_\sigma$, 
provided that the changed role of $f_0/\sigma$ is understood.
 
Each non-NG particle or resonance of mass $M_\mathrm{res}$ gives rise to
a pole factor which carries a linear combination $p = O(m_K)$ of the 
external momenta. The relevant coefficient is obtained from terms 
$\sim  p^4/M^2_\mathrm{res}$ in \emph{heavy-particle} expansions of these
pole factors
\begin{align}
p^4\bigl/\bigl(p^2 - M^2_\mathrm{res}\bigr) 
= p^2 + p^4/M^2_\mathrm{res} + \ldots \ \mbox{ for } M_\mathrm{res} \gg p 
\,.
\label{heavy}\end{align}
These expansions are nonchiral, i.e.\ they are not light-particle, 
small-momentum expansions of the type (\ref{irscale}). Evidently this 
technique assumes a clear scale separation between the NG and non-NG 
sectors.

Where does the $f_0(500)$ resonance fit into this scheme? Having it 
contribute as a light particle in chiral expansions and a heavy particle 
in Eq.~(\ref{heavy}) would be double counting.

In $\chi$PT$_3$, the answer is that the $f_0(500)$ does not belong to the 
NG sector, so it is treated as a heavy resonance. The obvious lack of 
scale separation with the $K,\eta$ NG bosons (Fig.~\ref{fig:goldstone})
makes this proposal unworkable.

In $\chi$PT$_\sigma$, the problem disappears because $f_0/\sigma$ 
is assigned to the NG sector. Its contributions are already taken into 
account in chiral expansions, so logically, it must be \emph{excluded} 
from the saturation procedure of \cite{Eck89}. That is in line with the
requirement that saturation be restricted to the \emph{non-NG sector}.
Scale separation of the NG and non-NG sectors works well for 
$\chi$PT$_\sigma$ (Fig.~\ref{fig:goldstone}), so the heavy-particle 
conditions $M_\mathrm{res} \gg p$ for $p = O(m_K)$ are satisfied.

In practice, $\chi$PT$_\sigma$ coefficients such as those in 
Eqs.~(\ref{invNLO}) and (\ref{d>4NLO}) are not easily evaluated, 
because the analysis requires data for soft $\sigma$ as well as 
soft $\pi,K,\eta$ amplitudes.

\section{Electromagnetic properties of mesons}
\label{Electromag}
In $\chi$PT$_\sigma$, the electromagnetic interactions of NG bosons 
are of great interest because
\begin{itemize}
\item The amplitudes for $K_S \to \gamma\gamma$ and $\gamma\gamma \to \pi\pi$ 
      can be used to analyse $K \to 2\pi$ (Sec.~\ref{weak_emag}).
\item The electromagnetic trace anomaly (\ref{eqn:em_anomaly}) and hence 
      the Drell-Yan ratio can be estimated at the infrared fixed point 
      $\alpha_s = \alpha^{}_\mathrm{IR}$.
\item In $\gamma\gamma$ channels, meson loops can produce Li-Pagels  
      singularities $\sim 1/m^2_{\pi ,K, \sigma}$ and hence amplitudes which 
      compete with $\sigma$-pole tree diagrams.
\end{itemize}

Photon interactions are introduced as in $\chi$PT$_3$, with the added 
requirement that the chiral singlet field $\sigma$ is gauge invariant.
So under local $U(1)$ transformations, we have
\begin{equation}
\sigma \to \sigma\,,\quad
U\to e^{-i\lambda(x)Q}Ue^{i\lambda(x)Q}\,,
\end{equation}
where $Q = \tfrac{1}{3}\mathrm{diag}(2,-1,-1)$ is the quark-charge 
matrix.  Gauge invariance can be satisfied minimally by introducing 
a covariant derivative for $U$,
\begin{equation}
D_\mu U = \del_\mu U +ieA_\mu [Q,U]\,,
\label{DelU}
\end{equation}
where $A_\mu$ is the photon field. However this is not sufficient:
it does not change the scaling properties of the effective Lagrangian, 
and so cannot produce an electromagnetic trace anomaly (\ref{eqn:em_anomaly})
proportional to $F_{\mu\nu}F^{\mu\nu}$.

The operator $F_{\mu\nu}F^{\mu\nu}$ has dimension 4, so we need an action
which, when varied, produces a scale \emph{invariant} result. This can 
happen only if the scaling property is \emph{inhomogeneous}. The 
$\sigma$ field has a scaling property (\ref{sigma_scale}) of that type, 
from which it is evident that the effective Lagrangian must contain a 
nonminimal term of the form
\begin{equation}
{\cal L}_{\sigma\gamma\gamma} = 
\tfrac{1}{4} g_{\sigma\gamma\gamma}\sigma F_{\mu\nu}F^{\mu\nu} \,.
\label{non-min}
\end{equation}  
This is the effective vertex first considered by Schwinger \cite{Schw51}
in his study of the gauge invariance of fermion triangle diagrams.

Originally, the electromagnetic trace anomaly (\ref{eqn:em_anomaly}) was
derived in the context of broken scale invariance (before QCD and asymptotic
freedom), so the ultraviolet limit defining the Drell-Yan ratio $R$ was 
nonperturbative.  A comparison of Eqs.~(\ref{eqn:em_anomaly}) and 
(\ref{non-min}) in the tree approximation, or equivalently $\sigma$-pole 
dominance of $\theta^\mu_\mu$ (PCDC), led to the conclusion \cite{RJC72,Ell72} 
that the coupling of $\sigma$ to $\gamma\gamma$ is proportional to $R$.

In the current context, there are two important modifications to this 
argument.

The first is to identify ``$R$'' correctly. In the physical region
$0 < \alpha_s < \alpha^{}_\mathrm{IR}$, asymptotic freedom controls the 
ultraviolet limit and produces a perturbative answer
\begin{equation}
R^{}_\mathrm{UV} = \sum\{\mbox{quark charges}\}^2 = 2\,, \ N_f = N_c = 3
\label{eqn:RUV}
\end{equation}
for $N_f = 3$ light flavors and $N_c = 3$ colors. However, the hard gluonic 
operator $G^2$ in $\theta^\mu_\mu$ prevents PCDC from being used to relate 
low-energy amplitudes to asymptotically free quantities like $R^{}_\mathrm{UV}$. 
Instead, in the lowest order of $\chi$PT$_\sigma$, we use amplitudes defined 
at the infrared fixed point where the gluonic trace anomaly vanishes and 
so PCDC can be tested.  At the infrared fixed point $\alpha_s 
= \alpha^{}_\mathrm{IR}$, there is no asymptotic freedom, so the UV limit 
of $e^+ e^- \to$ hadrons produces a \textit{nonperturbative} value 
$R_\mathrm{IR}$ which has to be determined theoretically.  Thus we expect 
$g_{\sigma\gamma\gamma}$ to be related to $R_\mathrm{IR}$.

The second modification is a surprise.  In $\gamma\gamma$ channels, meson-loop 
integrals produce inverse Li-Pagels singularities \mbox{$\sim M^{-1}$} in the 
chiral limit $M \sim 0$, where $M$ is the $\pi, K, \eta$ (mass)$^2$ matrix 
(\ref{mass_matrix}). These infrared singularities are strong enough to allow 
$\pi^\pm,K^\pm$ one-loop diagrams to have the {\it same} chiral order as tree 
amplitudes containing the anomalous vertex in (\ref{non-min}). This means that 
naive PCDC ($\sigma$-pole dominance) does not work when $\gamma\gamma$ channels 
are present; for example, the $\sigma \to \gamma\gamma$ coupling turns out 
to be proportional to $(R_\mathrm{IR} - 1/2)$, not $R_\mathrm{IR}$.  Similar 
problems are not encountered for PCAC, partly because loop corrections to 
PCAC are limited by the negative parity of the corresponding Nambu-Goldstone 
bosons.

It becomes less surprising when the power-counting rule (\ref{GLrule}) 
for electromagnetic corrections to $\chi$PT expansions is considered.

A standard treatment of $\chi$PT \cite{Gasser84,Gasser85} is to require that 
the effective Lagrangian be invariant under {\it local} chiral 
$SU(N_f)_L \times SU(N_f)_R$ transformations.  This requirement is satisfied 
minimally by replacing ordinary derivatives $\del_\mu$ acting on $U$ fields 
with covariant ones
\begin{equation}
D_\mu U = \del_\mu U - \tfrac{i}{2}(v_\mu+a_\mu) U 
          + \tfrac{i}{2} U(v_\mu-a_\mu) \,,
\label{eqn:cov D}
\end{equation}
where the gauge fields $v_\mu(x)$ and $a_\mu(x)$ transform inhomogeneously 
under the respective vector and axial-vector subgroups of 
$SU(N_f)_L \times SU(N_f)_R$.  In order to match the chiral counting 
$\del_\mu U = O(p)$ used by Weinberg \cite{Wei79} to study pure pion 
processes in $\chi$PT$_2$, Gasser and Leutwyler proposed the rule
\cite{Gasser84,Gasser85}
\begin{equation}
a_\mu \sim v_\mu = O(p)\,.
\label{chiral rule}
\end{equation}
For electromagnetic processes, this requires the photon field $A_\mu$ 
obtained from
\begin{equation}
v_\mu = -2eQA_\mu \qquad \mbox{and} \qquad a_\mu = 0,
\end{equation}
to be counted as $O(p)$. As a result, one-loop meson amplitudes which couple 
(say) $\sigma$ to any number of external photons are of the same chiral order, 
namely $O(p^4)$.

In $\chi$PT$_\sigma$, where the global symmetry group includes dilatations, 
chiral gauge invariance is not sufficient to determine the chiral order for 
nonminimal operators such as (\ref{non-min}).  In Appendix \ref{AppB}, we 
generalize the Gasser-Leutwyler analysis to cover such cases. As a result:
\begin{enumerate}
\item Both Eq.~(\ref{chiral rule}) and the rule $A_\mu = O(p)$ remain valid.
\item The operator (\ref{non-min}) gives rise to a $O(p^4)$ vertex amplitude
of the same chiral order as one-loop meson graphs 
for $\sigma \to \gamma\gamma$. 
\item In the presence of photons, $\chi$PT$_\sigma$ corrections to lowest-order 
tree and loop diagrams still converge: each additional loop is suppressed by a 
factor $\sim M\ln M$ or $M$.
\end{enumerate}

In this Section, we consider lowest-order amendments to PCDC for the amplitude 
$\langle\gamma\gamma|\widetilde{\theta}^\mu_\mu|\mathrm{vac}\rangle$.

Let $\gamma_i = \gamma(\epsilon_i,k_i)$ represent a photon with 
polarization $\epsilon_i$ and momentum $k_i$, and let $F(s)$ be the
form factor defined by
\begin{equation}
\langle\gamma_1,\gamma_2|\widetilde{\theta}_{\mu}^{\mu}(0)|\mathrm{vac}\rangle 
= (\epsilon_{1}\cdot\epsilon_{2} k_{1}\cdot k_{2} 
-\epsilon_{1}\cdot k_{2}\epsilon_{2}\cdot k_{1}) F(s) \,.
 \label{form}
\end{equation}
The electromagnetic trace anomaly concerns the value of this form factor
at $s=0$: 
\begin{equation}
F(0) = -\tfrac{1}{3}\pi\alpha \int d^4x\, d^4y\, x\cdot y\,
 T \langle  
J^{\beta}(x) J_{\beta}(0) \theta_{\mu}^{\mu}(y) \rangle_{\mathrm{vac}} \,.
\label{form'}\end{equation}
At the fixed point $\alpha_s = \alpha^{}_{\mathrm{IR}}$, we have a theory
of broken scale invariance, so the conditions of the derivations in
\cite{RJC72,Ell72} are satisfied. The leading short-distance behavior of
both $\langle J_\alpha J_\beta \theta_{\mu\nu}\rangle_{\mathrm{vac}}$
and $\langle J_\alpha J_\beta \rangle_{\mathrm{vac}}$ is conformal, with 
no anomalous dimensions because $J_\alpha$ and $\theta_{\mu\nu}$ are 
conserved, and the soft $d < 4$ trace $\theta^\mu_\mu$ ensures convergence
of Eq.~(\ref{form'}) at $x \sim y \sim 0$. Therefore, we can write
down an exact anomalous Ward identity%
\footnote{There is similar result for $0 < \alpha_s  < \alpha^{}_{\mathrm{IR}}$ 
which involves $R^{}_\mathrm{UV}$ but has no practical use. See Appendix 
\ref{AppC}.} 
\begin{equation}
F(0) = \frac{2R^{}_{\mathrm{IR}}\alpha}{3\pi}\,, \ 
 \alpha_s = \alpha^{}_{\mathrm{IR}}\,.
\label{eqn:rjcform}
\end{equation}

The calculation of the form factor $F(s)$ in $\chi$PT$_\sigma$ involves two 
classes of diagrams (Fig.\ \ref{fig:anomaly_loops}):
\begin{figure}[t]
\center \includegraphics[scale=0.34]{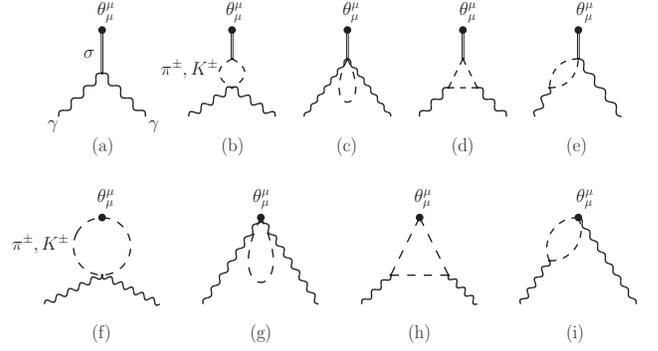}
\caption{Lowest order contributions to 
$\langle\gamma_1,\gamma_2|\widetilde{\theta}^\mu_\mu(0)|\mathrm{vac}\rangle$ 
in $\chi$PT$_\sigma$.  Diagram (a) represents the contact term proportional to 
$g_{\sigma\gamma\gamma}$, while diagrams (d), (e), (h), and (i) are 
each accompanied by an additional crossed amplitude (not shown).
Similar loop diagrams have been considered in $\chi$PT$_3$ for 
$K_S\to\gamma\gamma$ \cite{DAm86}, $K_L\to\pi^0\gamma\gamma$ \cite{Eck87}, 
and $\gamma\gamma\to\pi^0\pi^0$ \cite{Don88,Bij88}.
}
\label{fig:anomaly_loops}
\end{figure}
\begin{enumerate}
\item Dilaton pole diagrams (a-e) which produce a factorized amplitude
\begin{equation}
F_1(s) = 
{\cal A}_{\sigma\gamma\gamma} \frac{i}{s - m^2_\sigma} (-m_\sigma^2 F_\sigma) \,.
\label{F_1}
\end{equation}
Here ${\cal A_{\sigma\gamma\gamma}}$ includes a contact term 
$-ig_{\sigma\gamma\gamma}$ from diagram (a) and contributions from
one-loop  diagrams (b-e) with internal $\pi^\pm, K^\pm$ lines. 
\item A one-loop amplitude $F_2(s)$ from diagrams (f-i) with internal 
$\pi^\pm,K^\pm$ lines coupled to the vacuum via $\theta^\mu_\mu$.
\end{enumerate}

The $\sigma \to \gamma\gamma$ amplitude in Eq.~(\ref{F_1}) can be written 
\begin{equation}
{\cal A}_{\sigma\gamma\gamma} = - ig_{\sigma\gamma\gamma}
                             + \frac{i\alpha}{\pi F_{\sigma}} {\cal C}
       \sum_{\phi=\pi, K} m_{\phi}^{2}\Big(\frac{1+2I_{\phi}}{s}\Big) 
\label{eqn:Asig2gam}
\end{equation}
where the label $\phi = \pi^\pm \mbox{ or } K^\pm$ 
refers to the meson propagating around the loop in diagrams (b-e).
In Eq.~(\ref{eqn:Asig2gam}), the constant $\cal C$ is a combination of 
low energy coefficients
\begin{equation}
{\cal C} = 1 - \gamma_m - (1-c_1)\beta' 
\label{eqn:C}
\end{equation}
and $I_\phi$ is the relevant Feynman-parametric integral
\begin{align}
I_{\phi} = m_\phi^2\overset{\ \ 1}{\iint\limits_0}
 dz_{1}dz_{2}\,\theta(1-z_{1}-z_{2})\bigl/\bigl(z_{1}z_{2}s - m_{\phi}^2\bigr)
\label{eqn:feyn par}
\end{align}
for on-shell photons $k_1^2 = k_2^2 = 0$.  The constant $\cal C$ and integral 
$I_\phi$ also appear in the result for diagrams (f-i):
\begin{equation}
F_2(s) = \frac{\alpha}{\pi}({\cal C} - 2) 
       \sum_{\phi=\pi, K} m_{\phi}^{2}\Big(\frac{1+2I_{\phi}}{s}\Big)  \,. 
\end{equation}

The final step is to compare the answer for
\begin{equation}
F(s) = F_1(s) + F_2(s) 
\end{equation}
with the $s=0$ constraint (\ref{eqn:rjcform}). For that, we need the Taylor 
expansion
\begin{equation}
1+2I_{\phi} = - \frac{s}{12m_{\phi}^{2}} + O(s^{2}) \,.
\end{equation}
Summing the $\pi^{\pm}$ and $K^{\pm}$ contributions, we have 
\begin{equation}
\sum_{\phi=\pi, K} m_{\phi}^{2}\Big(\frac{1+2I_{\phi}}{s}\Big) 
= - \frac{1}{6} + O(s)  \,,
\label{Taylor}
\end{equation} 
and so find that the terms involving ${\cal C}$ cancel:
\begin{equation}
F(s) = g_{\sigma\gamma\gamma} F_\sigma + \alpha/3\pi + O(s) \,.
\end{equation}
Comparison with Eq.~(\ref{eqn:rjcform}) yields the desired relation%
 \footnote{The answer is simple because we chose a $\sigma$ field with 
 the scaling property (\ref{sigma_scale}). Constants like $\cal C$ 
 can appear if other definitions of $\sigma$ are used.\label{sigma_def}}
\begin{equation}
g_{\sigma\gamma\gamma} 
=  \frac{2\alpha}{3\pi F_\sigma} \Big(R^{}_{\mathrm{IR}} - \tfrac{1}{2} \Big)\,.
\label{eqn:gsig2gam}
\end{equation}
Evidently, the one-loop diagrams which produce the term $-\frac{1}{2}$ 
relative to $R^{}_{\mathrm{IR}}$ have the same chiral order as the tree diagram 
involving $g_{\sigma\gamma\gamma}$. This is an explicit demonstration of the 
way PCDC is modified by the inverse Li-Pagels singularities noted above for 
$\gamma\gamma$ channels.  

An estimate for $R^{}_{\mathrm{IR}}$ from Eq.~(\ref{eqn:gsig2gam}) is not 
straightforward because dispersive analyses of reactions such as 
$\gamma\gamma\to\pi\pi$ yield residues at the $f_0/\sigma$ pole 
proportional to the full amplitude 
${\cal A}_{\sigma\gamma\gamma}(s = m^2_\sigma)$ of Eq.~(\ref{eqn:Asig2gam}), 
not $g_{\sigma\gamma\gamma}$. Currently, we have no independent data about the
constant $\cal C$, apart from the weak constraint (\ref{inequal}) for 
$(1-c_1)\beta'$ and the inequality
\begin{equation}
-1 \leqslant 1 - \gamma_m < 2  
\end{equation}
from Eq.~(\ref{dim-mass}). We will argue below that numerically, 
these corrections are likely to be small compared with the electromagnetic 
trace anomaly. First, let us review what is known about $\gamma\gamma\to\pi\pi$ 
from dispersion theory.

The residue of the $f_0(500)$ pole was first extracted from the Crystal 
Ball data \cite{Xal90} by Pennington \cite{Pen06} and subsequently refined 
in several analyses \cite{Oll08,Mao09,Mou11,Hof11}. We use a recent 
determination \cite{Hof11} of the radiative width
\begin{equation}
\Gamma_{\sigma\gamma\gamma} = 2.0 \pm 0.2 \mbox{ keV} 
\label{eqn:Hof}
\end{equation} 
based on fits to data \cite{Gar10} of pion polarizabilities.%
\footnote{We do not use the alternative estimate 
\cite{Hof11} $\Gamma_{\sigma\gamma\gamma} = 1.7\pm 0.4$ keV
because it depends on scalar meson resonance saturation for low energy 
constants of $\chi$PT$_2$ expansions \cite{Gass05,Gass06} and also 
(tracing back via App.~D.2.2 of \cite{Bell94} to \cite{Amet92}) 
$\chi$PT$_3$ expansions. As noted in Sec.~\ref{strong} below 
Eq.~(\ref{heavy}), that places $f_0$ in the non-NG sector. It would 
be inconsistent for us to combine that with $\chi$PT$_\sigma$.}

In lowest order $\chi$PT$_\sigma$, the relevant diagrams for the 
process $\sigma\to\gamma\gamma$ are those shown in (a-e) of 
Fig.~\ref{fig:anomaly_loops}, but with $\sigma$ treated as an asymptotic
state. The narrow width approximation is valid in lowest order 
$\chi$PT$_\sigma$, so the magnitude of the full amplitude 
${\cal A}_{\sigma\gamma\gamma}$ at $s=m_\sigma^2$ is determined by 
\begin{equation}
\Gamma_{\sigma\gamma\gamma} 
= \frac{m_\sigma^3}{64\pi}|{\cal A}_{\sigma\gamma\gamma}|^2 \,. 
\label{eqn:Gam 2gam}
\end{equation}
Comparison with (\ref{eqn:Hof}) then gives
\begin{equation}
|{\cal A}_{\sigma\gamma\gamma}| = 0.068\pm 0.006 \ \mbox{GeV}^{-1} 
\label{eqn:modA}
\end{equation}
where the uncertainties have been added in quadrature.

The presence of lowest order meson loops in $\gamma\gamma$ channels 
implies that numerical results for the contact term depend on how the 
scalar field is defined.\footnotemark[14]
Consequently, care must be exercised when 
comparing our value with those found using $\chi$PT$_3$ or dispersion
theory --- definitions of ``the contact $f_0\gamma\gamma$ coupling'' 
are not necessarily equivalent. For example, the small values for these
couplings reported in dispersive analyses \cite{Ach07,Men08} could well 
be consistent with each other and with our result for the coupling 
${\cal L}_{\sigma\gamma\gamma}$ of Eq.~(\ref{non-min}).

In $\chi$PT$_\sigma$ we find that for $N_c$ large, it is the contact term 
which is the dominant contribution to ${\cal A}_{\sigma\gamma\gamma}$. This 
is because, relative to the single-quark loop diagrams associated with
$R^{}_\mathrm{IR} = O(N_c)$, terms from $\pi^\pm,K^\pm$ loop graphs involve 
an additional quark loop and so are suppressed by a factor $1/N_c$.  We 
therefore have
\begin{equation}
g_{\sigma\gamma\gamma} 
 = O\bigl(\sqrt{N_c}\bigr) \quad \mbox{and} \quad {\cal C} = O(1)
\end{equation}
in the large-$N_c$ limit and conclude%
 \footnote{This approximation is {\it not} required in our analysis of 
 $K_S\to\pi\pi$ in Sec.\ \ref{weak_emag}. Indeed $g_{\sigma\gamma\gamma}$
 does not appear anywhere. The key ingredient is the phenomenological 
 estimate (\ref{eqn:modA}) for the complete amplitude 
 ${\cal A}_{\sigma\gamma\gamma}$.}
\begin{equation}
{\cal A}_{\sigma\gamma\gamma}
  = -ig_{\sigma\gamma\gamma} + O\bigl(1/\sqrt{N_c}\bigr) \,.
\end{equation} 

From Eq.~(\ref{eqn:modA}) and within the large uncertainty due to that in 
$F_\sigma$, we estimate 
\begin{equation}
R^{}_{\mathrm{IR}} \approx 5\,. 
\end{equation}
This result is a feature of the nonperturbative theory at 
\begin{figure}[tb]
\center\includegraphics[scale=0.75]{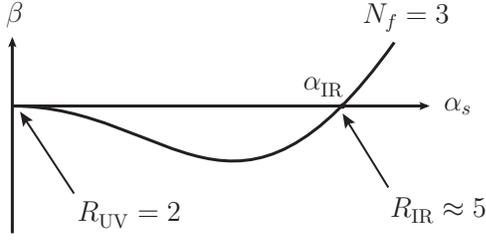}
\caption{Drell-Yan ratios $R^{}_{\mathrm{UV}}$ and $R^{}_{\mathrm{IR}}$ associated
 with the proposed $\beta/\psi$ function. For $e^+ e^- \to$ hadrons at high
 energies with $0 < \alpha_s < \alpha^{}_\mathrm{IR}$, the strong coupling 
 $\alpha_s$ runs to zero and the result $R^{}_{\mathrm{UV}}$ is perturbative 
 (asymptotic freedom). However if $\alpha_s$ is at $\alpha^{}_{\mathrm{IR}}$, 
 it cannot run, so we get a nonperturbative result $R^{}_{\mathrm{IR}}$ 
 associated with \emph{short}-distance scaling at the infrared fixed point.}  
\label{fig:beta_RIR}
\end{figure}%
$\alpha^{}_\mathrm{IR}$ (Fig.\ \ref{fig:beta_RIR}), so it has \emph{nothing} 
to do with asymptotic freedom or the free-field formula (\ref{eqn:RUV}).

\section{Weak interactions of mesons}
\label{weak_emag}
The most important feature of $\chi\mathrm{PT}_{\sigma}$ is that it
explains the empirical $\Delta I=1/2$ rule for nonleptonic kaon decays 
such as $K\rightarrow\pi\pi$. 

Problems explaining the data for nonleptonic kaon and hyperon decays were 
first recognised sixty years ago \cite{mgm54}. They became acute with the 
advent of three-flavor chiral perturbation theory. For $\chi$PT$_3$ applied to 
kaon decays, the dilemma is:
\begin{enumerate}
\item A fit to data in lowest nontrivial order, i.e.\ for $O(p^2)$ amplitudes 
      ${\cal A}_\mathrm{LO}$, would require the ratio of $\bm{8}$ to $\bm{27}$  
      couplings $|g_8/g_{27}|$ to be $\simeq 22$, much larger than any of the 
      reasonable estimates in the range (\ref{ratio}).
\item The main alternative is to accept Eq.~(\ref{ratio}) and argue that
      the dominant contribution comes from a NLO $O(p^4)$ term produced by 
      strong final-state interactions in the $0^{++}$ channel, e.g.\ via 
      a non-NG scalar boson $S$ \cite{Golo80, Volk88, Moro90, Pol02} for 
      which the pole diagram in Fig.\ \ref{fig:k_pipi} is $O(p^4/m_S^2)$, 
      with $m_S \not= 0$. Then the $\chi$PT$_3$ expansion diverges 
      uncontrollably,%
\footnote{The factor 22 is 70 times larger than the limit $\sim 0.3$ 
 prescribed by Eq.~(\ref{eqn:XPT3_fit}) for an acceptable fit.}%
      \begin{equation}
      \bigl|\mathrm{NLO}\bigl/\mathrm{LO}\bigr|_{\chi\mathrm{PT}_3} \simeq 22 
      \label{22}\end{equation}
      contradicting the premise that $\chi$PT$_3$ is applicable.
\end{enumerate}

Let us review option 1 in more detail. In the lowest order%
\footnote{Our aim is to solve the $\Delta I = 1/2$ puzzle \emph{without}
  using NLO terms. Weak NLO terms in $\chi$PT$_3$ \cite{weakNLO}, except 
  those depending on $f_0/\sigma$ (Sec.\ \ref{saturate}), become weak 
  NLO $\chi$PT$_\sigma$ terms when multiplied (as in Eq.~(\ref{d>4NLO}))
  by suitable powers of $e^{\sigma/F_\sigma}$. We expect these to produce 
  small corrections to our result.} 
of standard $\chi$PT$_3$, the effective weak Lagrangian
\begin{equation}
\left.\mathcal{L}_{\mathrm{weak}}\right|_{\sigma=0} 
= g_{8}Q_{8} + g_{27}Q_{27} + Q_{mw} + \mathrm{h.c.}\
\label{usual}\end{equation}
contains an octet operator \cite{Cro67}
\begin{equation}
Q_{8} = \mathcal{J}_{13}\mathcal{J}_{21} - \mathcal{J}_{23}\mathcal{J}_{11}
\ , \quad
\mathcal{J}_{ij} = (U\partial_{\mu}U^{\dagger})_{ij}
\label{eqn:octet}
\end{equation}
the $U$-spin triplet component \cite{Gaill74,RJC86} of a \textbf{27} operator
\begin{equation}
Q_{27} = \mathcal{J}_{13}\mathcal{J}_{21} 
           + \tfrac{3}{2} \mathcal{J}_{23}\mathcal{J}_{11}
\label{eqn:27-plet}
\end{equation}
and a weak mass operator \cite{Bern85}
\begin{equation}
Q_{mw} = \mathrm{Tr} (\lambda_6 - i\lambda_7)
      \bigl(g_MMU^\dagger + \bar{g}_MUM^\dagger\bigr) \,.
\label{eqn:weak mass}
\end{equation}
Although $Q_{mw}$ has isospin 1/2, it cannot be used to solve the 
$\Delta I = 1/2$ puzzle if dilatons are absent. 
When $Q_{mw}$ is combined with the strong mass term 
$\left.{\cal L}_\mathrm{mass}\right|_{\sigma = 0}$, it can be removed by 
a chiral rotation 
\begin{equation}
U\to \widetilde{U} = RUL^\dagger
\end{equation}
which aligns the vacuum such that 
\begin{equation}
\langle \widetilde{U}\rangle_\mathrm{vac} = I \ \mbox{ and } \
M = \mbox{real diagonal}.
\end{equation}
Therefore \cite{RJC86} $Q_{mw}$ has no effect on  $\chi$PT$_3$ 
low-energy theorems relating $K \to \pi\pi$ and $K \to \pi$
on shell, and so the conclusion that $|g_8/g_{27}|$ is unreasonably large 
($\approx$ 22) cannot be avoided.

In $\chi$PT$_\sigma$, the outcome is entirely different. 
First, we adjust the operator dimensions of $Q_8,\, Q_{27},$ and 
$Q_{mw}$ by powers of $e^{\sigma/F_\sigma}$
\begin{align}
{\cal L}_{\mathrm{weak}} &= Q_{8}\sum_n g_{8n}e^{(2 -\gamma_{8n})\sigma/F_\sigma}
+ g_{27}Q_{27}e^{(2-\gamma_{27})\sigma/F_\sigma}   \nonumber \\
&+ Q_{mw}e^{(3-\gamma_{mw})\sigma/F_\sigma} + \mbox{h.c.}\,,
\end{align}
as in Eqs.~(\ref{Lstr}) and (\ref{d>4NLO}) for the strong interactions,
with octet quark-gluon operators allowed to have
differing dimensions at $\alpha^{}_\mathrm{IR}$. The key point is that
the weak mass operator's dimension $(3-\gamma_{mw})$ bears no 
relation to the dimension $(3-\gamma_{m})$ of ${\cal L}_\mathrm{mass}$, 
so the $\sigma$ dependence of $Q_{mw} e^{(3-\gamma_{mw})/F_\sigma}$ 
cannot be eliminated by a chiral rotation. Instead, after aligning the vacuum, 
we find
\begin{align}
&\mathcal{L}^{\mathrm{align}}_{\mathrm{weak}}
 = \widetilde{Q}_{8}\sum_n g_{8n}e^{(2-\gamma_{8n})\sigma/F_\sigma}
 + g_{27}\widetilde{Q}_{27}e^{(2-\gamma_{27})\sigma/F_\sigma}   \nonumber \\
 &+ \widetilde{Q}_{mw}\bigl\{e^{(3-\gamma_{mw})\sigma/F_\sigma} 
 - e^{(3-\gamma_{m})\sigma/F_\sigma}\bigr\} + \mathrm{h.c.}\,,
\end{align}
where the tilde indicates that the $\bm 8$ and $\bm 27$ operators 
are now functions of the rotated field $\widetilde{U}$. As a result, 
there is a residual 
interaction $\mathcal{L}_{K_S\sigma} = g^{}_{K_S\sigma}K_{S}\sigma$
which mixes $K_{S}$ and $\sigma$ in \emph{lowest} $O(p^2)$ order  
\begin{equation}
g^{}_{K_S\sigma} 
 = (\gamma_{m} - \gamma_{mw})\mathrm{Re}\{(2m^2_K - m^2_\pi)\bar{g}_M
                    - m^2_\pi g_M\}F_{\pi}/2F_{\sigma} 
\end{equation}
and produces the $\Delta I = 1/2$ amplitude $A_{\sigma\textrm{-pole}}$
of Fig.~\ref{fig:k_pipi}.

At this point, we could simply choose $g_{K_S\sigma}$ to fit the rate 
for $K_S \to \pi\pi$, knowing that inserting the full $K_S \to \pi\pi$ 
amplitude into the standard loop calculation for $K_S \to \gamma\gamma$ 
\cite{DAm86} would give agreement with experiment. That would leave 
unclear what version of chiral perturbation theory in 
Fig.~\ref{fig:goldstone} is being used to analyse 
$K_S \to \gamma\gamma$.

So instead, we first apply $\chi$PT$_\sigma$ to $K_S \to \gamma\gamma$ 
and $\gamma\gamma\rightarrow\pi\pi$ in order to determine $g_{K_S\sigma}$, 
and then show that this gives a result for $K_S \to \pi\pi$ which agrees 
with experiment.

\begin{figure}[tb]
\center\includegraphics[scale=.43]{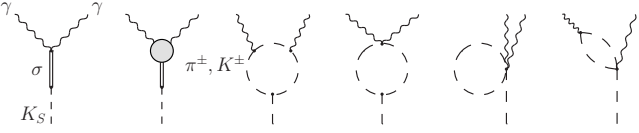}
\caption{\label{fig:loops} Lowest order diagrams for $K_{S}\to\gamma\gamma$ 
in $\chi\mathrm{PT}_{\sigma}$, including finite loop graphs \cite{DAm86}. 
The grey vertex contains $\pi^\pm,\,K^\pm$ 
loops as in the four $\chi$PT$_3$ diagrams to the right.  An analogous 
set of diagrams contributes to $\gamma\gamma \to\pi^{0}\pi^{0}$.}
\end{figure}

The scalar part ${\cal A}_{K\gamma\gamma}$ of the $K_{S}\rightarrow\gamma\gamma$ 
amplitude
\begin{equation}
{\cal A}_{\mu\nu} 
 = (g_{\mu\nu}k_1\cdot k_2 - k_{2\mu}k_{1\nu}){\cal A}_{K\gamma\gamma}
\end{equation}
receives three contributions at lowest order 
(Fig.~\ref{fig:loops})
\begin{equation}
{\cal A}_{K\gamma\gamma} = {\cal A}^{\mathrm{tree}}_\sigma 
+ {\cal A}^{\mathrm{loop}}_\sigma 
+ {\cal A}^{\mathrm{loop}}_{\pi, K}\,.
\end{equation}
The explicit expressions are
\begin{align}
&{\cal A}^{\mathrm{tree}}_\sigma + {\cal A}^{\mathrm{loop}}_\sigma = 
-2g^{}_{K_S\sigma}{\cal A}_{\sigma\gamma\gamma}\bigl/
    \bigl(m_K^2-m_\sigma^2\bigr)
\,, 
\nonumber \\
&{\cal A}^{\mathrm{loop}}_{\pi,K} = -2\frac{\alpha}{\pi F_\pi^3} 
 \bigl(g_8+g_{27}\bigr)
\sum_{\phi=\pi,K}(m_K^2-m_\phi^2) \Big(\frac{1+2I_\phi}{m_K^2}\Big)
\,, \label{eqn:kto2gam}
\end{align}
where the magnitude of ${\cal A}_{\sigma\gamma\gamma}$ is determined from 
Eq.~(\ref{eqn:modA}) and $I_\phi$ is the integral given by 
Eq.~(\ref{eqn:feyn par}).  If we neglect the $g_8$ and $g_{27}$ terms, 
we find 
\begin{equation}
|g^{}_{K_S\sigma}| \approx 4.4\times 10^{3}\,\mathrm{keV}^{2}
\label{gksig} \end{equation} 
to a precision $\lesssim 30\%$ expected for a three-flavor chiral 
expansion.

Now consider $K_S\to\pi\pi$ (Fig.\ \ref{fig:k_pipi}).  Eq.~(\ref{gksig}) 
and data for the $f_0$ width (Eq.~(\ref{numbers})) imply that the 
$\sigma$-pole diagram contributes (very roughly, 
given\footnotemark[12] $\sigma/f_0$'s width and near 
degeneracy with $K$)
\begin{equation}
\left|A_{\sigma\textrm{-pole}}\right| =   
\left|\frac{-ig^{}_{K_S\sigma}g_{\sigma\pi\pi}}{m_K^2 - m_\sigma^2}\right|
\approx 0.34\,\mathrm{keV}
\label{pole}
\end{equation}
to the full $I = 0$ amplitude%
\footnote{Our convention for the $K\to\pi\pi$ isospin amplitudes is that
 given in \cite{MRR}.}
\begin{equation}
{A}_{0} = \frac{\sqrt{3}}{F_\pi^3}\bigl(g_8+\tfrac{1}{6}g_{27}\bigr)
\bigl(m_K^2-m_\pi^2\bigr) + A_{\sigma\textrm{-pole}} \,.
\end{equation}
If the $g_{8,27}$ contributions are again neglected, 
\begin{equation}
\left|A_0\right| \simeq \left|A_{\sigma\textrm{-pole}}\right|
\end{equation}
we see that Eq.~(\ref{pole}) accounts for the large magnitude
of $A_0$ \cite{PDG}:
\begin{equation}
|A_{0}|_{\mathrm{expt.}} = 0.33\,\mathrm{keV} \,.
\end{equation}
We conclude that the observed ratio $|A_0/A_2| \simeq 22$ is mostly due to the 
dilaton-pole diagram of Fig.~\ref{fig:k_pipi}, that $g_{8} = \sum_n g_{8n}$ 
and $g_{27}$ have roughly similar magnitudes as simple calculations 
\cite{Feyn65,Feyn71,Gaill74,Alta74} indicate, and that only $g_{27}$ can 
be fixed precisely (from $K^{+}\rightarrow\pi^{+}\pi^{0}$). 

Consequently, the lowest $O(p^2)$ order of $\chi$PT$_\sigma$ solves the 
$\Delta I = 1/2$ problem for kaon decays. The chiral low-energy theorems 
which relate the on-shell%
\footnote{Ref.~\cite{RJC86} followed standard practice in current algebra. 
 It related the amplitude for $K \to \pi\pi$, where ${\cal H}_\mathrm{weak}$ 
 carries zero 4-momentum, to $K \to \pi$ for \emph{on-shell} kaons and pions, 
 where the relevant operator $[F_5,{\cal H}_\mathrm{weak}]$ obviously carries 
 nonzero 4-momentum. Ref.~\cite{RJC86} is often misquoted by authors who 
 implicitly set the momentum transfer for $K \to \pi$ equal to zero. 
 In Eq.~(\ref{tadpole}), $|\mathrm{vac}\rangle$ refers to the unique state 
 with translation invariance, so ${\cal H}_\mathrm{weak}$ carries momenta 
 whose square is $m_K^2$.}
$K \to 2\pi$ and $K \to \pi$ amplitudes have 
extra terms due to $\sigma$ poles, but the no-tadpoles theorem \cite{RJC86}
is still valid:
\begin{equation}
\langle K | \mathcal{H}_{\mathrm{weak}} |\mathrm{vac}\rangle
= O\bigl(m_s^2 - m_d^2\bigr)\,, \ K\mbox{ on shell}.
\label{tadpole}
\end{equation}

\section{Remarks}

Why must the $0^{++}$ particle be a dilaton in order to explain the 
$\Delta I = 1/2$ puzzle for $K$ decays?  Because the property 
$m_\sigma \to 0$ in the chiral-scale limit (\ref{scale})
is essential.  As is evident from Eq.~(\ref{22}), assuming scalar 
dominance by a non-NG particle contradicts the basic premise of chiral 
theory that at low energies, the NG sector dominates the non-NG sector. 
That is why none of the authors proposing scalar dominance by a non-NG 
particle since 1980 \cite{Golo80} claimed to have solved the puzzle or 
persuaded others to stop working on other proposals, such as penguin 
diagrams \cite{ITEP}, the large-$N_c$ limit \cite{largeN,Bur14}, or 
QCD sum rules \cite{QCDsum}.

\begin{figure}[t]
   \centering\includegraphics[scale=0.42]{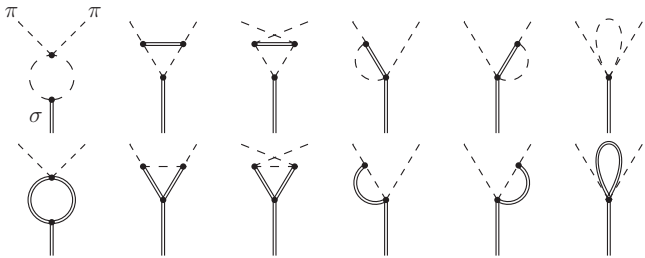}
   \caption{Next to lowest order (NLO) diagrams which contribute to the 
   resonance structure of $f_0/\sigma$ in $\chi$PT$_\sigma$.  Ultraviolet 
   divergences arising from the loops are absorbed (in the usual way) by 
   counterterms in the NLO Lagrangian.}
   \label{fig:width}
\end{figure}

Our resolution of the $\Delta I = 1/2$ puzzle is distinguished by not
being \emph{ad hoc}. It is part of a wider program to obtain numerically 
convergent three-flavor chiral expansions for amplitudes involving 
$0^{++}$ channels, i.e.\ where $\chi$PT$_3$ clearly fails (Sec.~\ref{Motiv}). 
So far, we can say only that lowest order $\chi$PT$_\sigma$ appears to 
be a good approximation. More stringent tests of convergence have yet 
to be developed because loop corrections involve couplings like 
$\sigma\sigma\pi\pi$ for which we lack data. %
An important example is the shape of the $\sigma$ resonance
at NLO (Fig.\ \ref{fig:width}), where the higher order corrections to 
(\ref{width}) have yet to be determined. This will require explicit 
calculations which include numerical fits for the $\sigma\sigma\sigma,
\sigma\sigma\pi\pi,\ldots$ couplings.

Another test could be to invent a unitarization procedure for 
$\chi$PT$_\sigma$ and check whether (unlike $\chi$PT$_3$) it produces 
\emph{small} corrections to lowest order results. 

The basis of our work on approximate scale and chiral $SU(3)_L\times SU(3)_R$ 
symmetry in QCD should be carefully distinguished from what is postulated 
in analyses of walking gauge theories. As noted by Del Debbio \cite{Del10},
in such theories, ``the infrared fixed point \ldots describes the physical 
properties of theories which are scale invariant at large distances, where 
the field correlators have power-like behaviours characterized by the 
anomalous dimensions of the fields."  That means that there is no mass
gap at the fixed point: scale condensates are assumed to be absent.

Our view of physics at the infrared fixed point is quite different. The 
Hamiltonian becomes scale invariant for massless $u,d,s$ quarks, but the 
vacuum is \emph{not} scale invariant because of the condensate 
$\langle \bar{q}q \rangle_{\mathrm{vac}} \not= 0$. It sets the scale 
of the mass gap for hadrons $\rho, \omega, N, \eta', \ldots$ in the 
non-NG sector (Sec.~\ref{Motiv} below Eq.~(\ref{eqn:phase})). 
For example, at the infrared fixed point in Fig.\ \ref{fig:beta_RIR}, 
$e^+e^-\to$ hadrons at low or intermediate energies has thresholds and 
resonances similar to QCD at similar energies.  Scaling behaviour sets in 
only at {\it high} energies.\footnotemark[3]

A result of this fundamental difference is that our hypothesis of an
infrared fixed point for $N_f = 3$ is not tested by lattice 
investigations done in the context of walking gauge theories.
Those investigations are based on criteria like Miransky scaling 
\cite{Miransky} which assume that a theory cannot have an infrared 
fixed point if it does not display the behavior described above in the
quote from Del Debbio.

More generally, our view is that theoretical evidence for or against our
proposal in Fig.~\ref{fig:beta} is inconclusive. Various definitions
of the QCD running coupling can be be readily compared in low orders of
perturbation theory, but it is not at all clear which definitions are 
physically equivalent beyond that. The key nonperturbative requirements 
for a running coupling are that its dependence on the magnitude of a 
space-like momentum variable be monotonic and analytic. Gell-Mann and 
Low \cite{GML} achieved this for QED, but these properties are hard to
establish for QCD running couplings. A lack of equivalence of these
definitions may explain why differing results for infrared fixed points
are obtained.

Unfortunately, our analysis does not explain the failure of chiral theory 
to account for nonleptonic $|\Delta S| = 1$ hyperon decays. We have 
shown that octet dominance is not necessary for $K$-decays,  
but that makes no difference for hyperon decays: the Pati-Woo 
$\Delta I = 1/2$ mechanism \cite{Pati71} forbids all contributions from 
\textbf{27} operators.

\begin{acknowledgments}
We thank Ross Young, Peter Minkowski, Martin Hoferichter, and Gilberto 
Colangelo for valuable discussions at various stages of this work.  LCT 
thanks Mary~K.~Gaillard and her colleagues for their kind hospitality 
at the Berkeley Center for Theoretical Physics, where part of this work 
was completed.  LCT is supported in part by the Australian-American 
Fulbright Commission, the Australian Research Council, the U.S. 
Department of Energy under Contract DE-AC02-05CH11231, the National 
Science Foundation under grant PHY-0457315, and the Federal Commission 
for Scholarships for Foreign Students (FCS).  The Albert Einstein Center 
for Fundamental Physics at the University of Bern is supported by the 
``Innovations- und Kooperationsprojekt C-13" of the ``Schweizerische 
Universit\"{a}tskonferenz SUK/CRUS".
\end{acknowledgments}

\emph{Note Added.} A similar chiral Lagrangian with a technicolor 
``dilaton'' has just been published by Matsuzaki and Yamawaki \cite{Mat13}. 
They acknowledge our prior work \cite{us}, but 
say that they do not believe it to be valid for hadronic physics. 
The basis for this assertion is the claim (footnote 8 of \cite{Bando}) 
that ``light'' dilatons are forbidden by the one-loop formula 
$-(6\pi)^{-1}(33 - 2N_f)\alpha_s^2$ for the QCD $\beta$ function.
The problems with this are that (a) the relevant limit is infrared, 
not ultraviolet, and (b) for $\alpha_s$ large, the one-loop formula
violates the analyticity bound \cite{Kras81} 
$\beta \gtrsim -\alpha_s\ln\alpha_s$.

\appendix
\section{Chiral Order in $\bm{\chi}$PT$_\sigma$}
\label{AppA}

In 1979, Weinberg \cite{Wei79} found that successive terms in the $\chi$PT$_2$ 
expansion of amplitudes with pionic external and internal lines and external 
momenta $p \sim m_\pi$ obey the ``power counting'' rule that each additional 
loop produces a factor $\sim p^2$ or, if there is a Li-Pagels singularity
\cite{LiPag,Pagels75}, $p^2 \ln p$. With essentially no change in the 
analysis, this rule can be extended to $\chi$PT$_3$ for pure $\pi, K, \eta$ 
amplitudes with $p \sim m_\pi, m_K, m_\eta$. Here we extend Weinberg's 
rule to $\chi$PT$_\sigma$ for amplitudes with internal and external lines 
restricted to the corresponding NG sector $\pi, K, \eta, \sigma$ 
(Fig.~\ref{fig:goldstone}).

This extension is not entirely straightforward because $\chi$PT$_\sigma$ is
really the special case
\begin{equation}
\alpha_s - \alpha^{}_\mathrm{IR} =O(M \mbox{ or } \del\del)\ \mbox{ for }\
 \del\del = O(M)   
\label{expand}
\end{equation}
of a double expansion\footnotemark[10] in the quark mass matrix 
$M$ about zero and the running coupling $\alpha_s$ about $\alpha^{}_\mathrm{IR}$.
In higher orders, critical exponents such as $\beta'(\alpha^{}_\mathrm{IR})$
and $\gamma_m(\alpha^{}_\mathrm{IR})$ become leading terms of expansions in 
$M$. For example, when $\gamma^{}_{G^2}(\alpha_s)$ in Eq.~(\ref{gamma}) is 
expanded about the fixed point, the dimension (\ref{dim-anom}) of the 
gluonic anomaly is corrected as follows:
\begin{equation}
d_\mathrm{anom} \to 4 + \beta'(\alpha^{}_\mathrm{IR}) 
  + \bigl(\alpha_s - \alpha^{}_\mathrm{IR}\bigr)\gamma'_{G^2}(\alpha^{}_\mathrm{IR})
  + \ldots
\end{equation}
Then terms in the $\chi$PT$_\sigma$ Lagrangian will include corrections of 
the form
\begin{equation}
e^{\beta'\sigma/F_\sigma} \to e^{\beta'\sigma/F_\sigma}\big\{1 + {\cal O}\bigr\}
\end{equation}
where ${\cal O}$ accounts for the effects of powers of the QCD factor 
$\alpha_s - \alpha^{}_{\mathrm{IR}}$. In the effective theory, this factor
may correspond to either an explicit $M$ factor or two derivatives 
$\del\ldots\del$ not necessarily acting on the same field. A power 
$(\alpha_s - \alpha^{}_{\mathrm{IR}})^p$ will then produce a superposition
of terms
\begin{equation}
\sim \{\mbox{$2k\ \del$ operators on up to $2k$ fields}\} M^{p-k} \,.
\end{equation}
Therefore ${\cal O}$ is in general an operator depending on powers of 
$\sigma$, $M$ and $\del\del$.

So in a higher chiral order, terms in the effective Lagrangian may  
involve $\sigma$-dependent factors
\begin{equation}
\sigma^{\mathrm{integer}\ >\ 0}\exp\bigl(\{\mathrm{constant}\}\sigma/F_\sigma\bigr)
\end{equation}
which do \emph{not} scale homogenously under the transformations 
(\ref{sigma_scale}). Indeed, terms of that type appear as renormalization 
counterterms for $\chi$PT$_\sigma$ loop expansions. The proliferation of
low-energy coupling constants due to inhomogenous scaling, with constraints 
between them possible but not obvious, makes the phenomenology of 
higher-order $\chi$PT$_\sigma$ challenging.

Fortunately, these complications do not impede the extension of Weinberg's
rule to $\chi$PT$_\sigma$. Our approach resembles Sec.\ 3.4.9 of the
review \cite{Scherer12}. 

Let $\phi$ refer to the spin $0^-$ octet $\pi,K,\eta$. In momentum space, a 
general vertex involving $\sigma$ and $\phi$ fields produces terms
\begin{equation}
\sim p_v^d m_\phi^{2k}m_\sigma^{2k'} \ , \quad\mbox{integers } d,k, k'
\label{eqn:vertex}
\end{equation}
where $p_v$ refers to components of the various vertex momenta and 
the product $p_v^d$ has degree $d$ when all $p_v$ are scaled to $tp_v$. 
The aim is to determine the behavior of Feynman diagrams under the 
rescaling 
\begin{equation}
p_e \to tp_e \,, \qquad m_\phi^2\to t^2m_\phi^2 \,, 
 \qquad m_\sigma^2 \to t^2m_\sigma^2
\label{eqn:scaling}
\end{equation}
of all external momenta $p_e$ and masses $m_{\phi,\sigma}$ 
of the NG bosons $\phi$ and $\sigma$. Note that the dilaton mass 
$m_\sigma$ scales in the same way as $m_\phi$, in keeping with the
discussion\footnotemark[10] 
of Eqs.~(\ref{dilaton-mass}), (\ref{L_anom}) and (\ref{expand}).

Let ${\cal A}(p_e,m_\phi,m_\sigma)$ be a connected amplitude given 
by a diagram with $I_{\phi}$ internal $\phi$ lines, $I_\sigma$ internal 
$\sigma$ lines, and $N_{dkk'}$ vertices of the form (\ref{eqn:vertex}). 
External lines are amputated (e.g.\ placed on shell) and the factor 
$\delta^4(\sum p_e)$ for momentum conservation is omitted. Apart from 
Li-Pagels logarithms \cite{LiPag,Pagels75} produced by loop integrals, 
each internal NG boson line
\begin{equation}
 \int\!\! \frac{d^4k}{(2\pi)^4}\frac{i}{k^2-m_{\phi,\sigma}^2 + i\epsilon}
\end{equation}
contributes a factor $t^2$ under (\ref{eqn:scaling}), 
so $\cal A$ scales with a chiral dimension or order given by
\begin{equation}
D = 4 + 2I_\phi  + 2I_\sigma +  \sum_{d,m,m'} N_{dmm'}(d+2m+2m'-4)\,.
\label{eqn:partial D}
\end{equation}
The number of independent loops $N_\ell$ for a graph is related to 
the total number of vertices $V=\sum_{dmm'} N_{dmm'}$ by the geometric 
identity
\begin{equation}
N_\ell = I_\phi + I_\sigma - V + 1\,.
\end{equation}
Substituting this identity into (\ref{eqn:partial D}) gives a result
\begin{equation}
D = 2 +  \sum_{d,m,m'} N_{dmm'}(d+2m+2m'-2) + 2N_\ell\,,
\label{eqn:chiral D}
\end{equation}
similar to that obtained originally \cite{Wei79} by Weinberg. 

The feature of Eq.\ (\ref{eqn:chiral D}) worth emphasizing is that the
loop number $N_\ell$ places a lower bound on $D$. That is because the 
vertex contribution (\ref{eqn:vertex}) must have chiral dimension 
$\geqslant 2$, i.e.
\begin{equation}
d+2m+2m'-2 \geqslant 0
\end{equation}
and so (\ref{eqn:chiral D}) implies the general inequality
\begin{equation}
D \geqslant 2 + 2N_\ell \,.
\label{loops}
\end{equation}
The case $D=2$ includes and is limited to tree graphs produced in lowest 
order, i.e.\ by vertices (\ref{eqn:vertex}) with chiral dimension $2$.

Given that the Li-Pagels infrared singularity for $N_\ell$-loop diagrams 
is $O(\ln^\ell t)$ at most, we see that each new loop is suppressed 
by a factor of at most $t^2\ln t$ for small $t$, as in $\chi$PT$_2$ and 
$\chi$PT$_3$. The extra logarithm is too weak to allow a given loop diagram 
to compete with diagrams with a smaller number of loops. In particular,
both $\sigma$-pole dominance (PCDC) and $\phi$-pole dominance (PCAC) are
valid for pure NG processes in lowest order. This is consistent with the 
discussion of the $\sigma$ width in Sec.\ \ref{strong}.

A further result is that higher-order versions of the constraint 
(\ref{L_anom}) on ${\cal L}_\mathrm{anom}$ are not needed. Once imposed,
it can be maintained to any order.

Apart from remarks about the $\sigma NN$ coupling in Sec.\ 
\ref{Motiv}, the analysis in this paper is limited to the NG sector.
Chiral power counting in the presence of baryons and other non-NG particles 
is a nontrivial matter even for ordinary $\chi$PT \cite{Scherer12}.

\section{External Currents and Wilson Operators in $\bm{\chi}$PT}
\label{AppB}

This Appendix concerns the effect of operators such as the lowest order 
$(\bm{8},\bm{1})$ and  $(\bm{1},\bm{8})$ currents
\begin{equation}
{\cal F}^{}_\mu = F_\pi^2  e^{2\sigma/F_\sigma} 
U i\del_\mu U^\dagger\
\mbox{ and }\
{\cal F}^\dagger_\mu = F_\pi^2  e^{2\sigma/F_\sigma} 
U^\dagger i\del_\mu U 
\label{currents}
\end{equation} 
on chiral power counting in the NG sector. This arises from the discussion in 
Sects.~\ref{Intro} and \ref{Electromag} of the Gasser-Leutwyler rule 
(\ref{GLrule}) and the failure of naive $\sigma$-pole dominance (PCDC) 
for $\sigma \to \gamma\gamma$, where loop diagrams with inverse-power 
Li-Pagels singularities compete with the tree diagram. These singular
powers are counted automatically if Appendix \ref{AppA} is extended to 
include vertices due to external operators carrying low momenta.

Vertices of the currents (\ref{currents}) have chiral order 1 because of 
the presence of a single derivative $\del = O(p)$. At first sight, 
counting a single power for the corresponding current sources (\ref{GLrule})
and (\ref{chiral rule}) seems obvious. For (say) a photon insertion 
in an internal propagator, we have
\begin{equation}
O\bigl(1/p^2\bigr) \longrightarrow O\bigl(A\times p/p^4\bigr)
\end{equation} 
with no change in loop number, so the chiral order for amplitudes with
photons can be matched to those without by choosing $A_\mu \sim p$. 

What is less obvious is the idea that these rules remain valid for sources
of currents in QCD itself, where they enter \emph{linearly} in the action,
e.g.
\begin{equation}
S^{}_{\mathrm{QCD}} \longrightarrow 
  S^{}_{\mathrm{QCD}} + \int d^4x A_\mu \bar{q}\gamma^\mu Q q 
\end{equation}
but give rise to nonlinear polynomial dependence in the effective theory.
For example, in addition to terms linear in $A_\mu$, the effective theory
contains anomalous terms proportional to $F^{\mu\nu}\widetilde{F}_{\mu\nu}$ 
for $\pi^0 \to \gamma\gamma$ and $F^{\mu\nu}F_{\mu\nu}$ for
$\sigma \to \gamma\gamma$, as well as non-anomalous powers of $A_\mu$
permitted by electromagnetic gauge invariance. Why should the rule 
$A_\mu \sim p$ for linear terms in the effective theory also hold
for terms nonlinear in $A_\mu$?

The reason is that infrared powers in NG-meson loop integrals are generated 
by a single mass scale: $M$. Therefore chiral order can be inferred from ordinary 
(non-operator) dimensionality. From $A_\mu \sim (\mbox{length})^{-1}$, we 
can conclude e.g.
\begin{equation}
F^{\mu\nu}F_{\mu\nu} = O(p^4) \,.
\end{equation}
The extra two derivatives in $F^2$ compared with $A_\mu A_\nu$ are 
responsible for the failure of the $\sigma$ vertex
(\ref{non-min}) to dominate one-loop meson contributions to 
$\sigma \to\gamma\gamma$.

When the lowest chiral order for a process induced by external currents 
mixes tree and loop diagrams, the rules of Appendix \ref{AppA} must be
amended. First, diagrams formed entirely from current vertices, i.e.\ with 
no vertices of the relevant $\chi$PT Lagrangian for strong interactions,
should be classified according to their loop number: tree, one-loop,
and sometimes higher. Then for each class, adding an internal loop 
produced by strong-interaction vertices increases the chiral order by at 
least 2. So the mixing of loop numbers for a given chiral order persists in 
higher orders, but the overall convergence rule that each new internal loop 
is suppressed by $M^2\ln M$ or $M^2$ is maintained.

Evidently, gauge invariance and the restriction to currents as external
operators is not essential for this discussion. All we need is a source 
${\cal J}$ of unique (non-operator) dimensionality for a QCD Wilson 
operator. This will generate a polynomial in ${\cal J}$ for the effective
theory with a chiral-order rule for ${\cal J}$. For example, let $\cal S$ 
and $\cal P$ be sources for $\bar{q}q$ and $\bar{q}\gamma_5q$ in QCD 
corresponding to
\begin{equation}
\bigl\{U \pm U^\dagger\bigr\}e^{(3 - \gamma_m)\sigma}
\end{equation} 
in lowest-order $\chi$PT$_\sigma$. Then insertion of this operator into
a $\phi$ or $\sigma$ propagator yields
\begin{equation}
O\bigl(1/p^2\bigr) \longrightarrow 
O\bigl(\{{\cal S}\mbox{ or }{\cal P}\}/p^4\bigr)
\end{equation} 
so keeping the chiral order constant, we rediscover the rule \cite{Gasser84}
\begin{equation}
{\cal S}\sim {\cal P} \sim O(p^2)\,.
\end{equation}
For a Wilson operator represented by a lowest order $\chi$PT operator whose
vertices are of chiral order $k$, the result is
\begin{equation}
{\cal J} \sim O(p^{2-k})\,.
\end{equation}

\section{Electromagnetic Trace Anomaly in QCD}
\label{AppC}

Originally, before QCD was invented, the electromagnetic trace anomaly 
was derived \cite{RJC72,Ell72} assuming a theory of broken scale 
invariance for strong interactions \cite{Wilson69}:
\begin{equation}
\mbox{dim } \theta^\mu_\mu < 4\,.
\label{broken}
\end{equation}
This anomaly relates the amplitude 
$T\langle \theta^\mu_\mu J_\alpha J_\beta \rangle_\mathrm{vac}$ in
the zero-energy limit (\ref{form'}) to the Drell-Yan ratio for the three-flavor 
theory at infinite%
\footnote{Note that the result (\ref{eqn:rjcform}) is exact; it is 
\emph{not} due to an estimate at some large but finite energy.  For 
example, it does not assume duality \cite{duality}.}
center-of-mass energy. Our application (\ref{eqn:rjcform}) is
restricted to $\alpha_s$ being exactly at the fixed point 
$\alpha^{}_{\mathrm{IR}}$, where broken scale invariance is still 
valid.

In the physical region $0 < \alpha_s  < \alpha^{}_{\mathrm{IR}}$ of
QCD, broken scale invariance, with its anomalous power laws in the 
ultraviolet limit for all operators except conserved currents, is
not valid, because the gluonic trace anomaly in Eq.~(\ref{eqn:anomaly}) 
violates Eq.~(\ref{broken}). However the ultraviolet requirements
of the derivation can be checked directly by using asymptotic freedom:
the leading short-distance behaviors of 
\begin{equation}
T\langle J_\alpha J_\beta \theta_{\mu\nu} \rangle_\mathrm{vac}\ , \
T\langle J_\alpha J_\beta \theta_{\mu\nu} \rangle_\mathrm{vac}\ 
\mbox{ and }\
T\langle J_\alpha J_\beta \theta^\mu_\mu \rangle_\mathrm{vac}
\label{three}
\end{equation}
are given by one-loop amplitudes with massless propagators, which
is a special case of what was assumed for broken scale invariance.
(The last amplitude in (\ref{three}) is needed to ensure convergence 
of Eq.~(\ref{form'}) at $x \sim y \sim 0$.) The fact that some 
nonleading terms die off as inverse logarithms instead of inverse powers 
has no effect on the derivation.

So we conclude that the derivation can also be carried through for 
QCD in the physical region, despite the hard breaking of scale 
invariance by the gluonic term in $\theta^\mu_\mu$. The result is
an exact relation
\begin{equation}
F(0) = \frac{2 R^{}_\mathrm{UV} \alpha}{3\pi} \ 
\mbox{ for } \ 0 < \alpha_s  < \alpha^{}_{\mathrm{IR}}
\label{QCDform}
\end{equation}
in terms of the perturbative ratio $R^{}_\mathrm{UV} = 2$ of 
Eq.~(\ref{eqn:RUV}).

Comparing Eqs.~(\ref{eqn:rjcform}) and (\ref{QCDform}), we see that 
$F(0)$ is discontinuous in $\alpha_s$ at $\alpha^{}_{\mathrm{IR}}$. That 
is not a problem because the $\sigma$ pole and charged $\pi^\pm, K^\pm$ 
loops can produce singular behaviour such as 
\begin{equation}
\sim \frac{q^2}{m_\sigma^2 - q^2} \ \mbox{ for } q, m_\sigma \sim 0 \,.
\end{equation}
However a relation between $R^{}_\mathrm{UV}$ and $R^{}_{\mathrm{IR}}$ 
\emph{cannot} be established because the condition (\ref{L_anom}) 
for ${\cal L}_\mathrm{anom}$ is not valid for an expansion not 
about $\alpha^{}_{\mathrm{IR}}$. 

\clearpage

\end{document}